\documentclass[pra,twocolumn,longbibliography]{revtex4-1}
\usepackage{graphicx}  
\usepackage{dcolumn}   
\usepackage{bm}        
\usepackage{verbatim}   
\usepackage{graphicx}
\usepackage{epstopdf}
\usepackage{footnote}
\usepackage{epsfig}
\usepackage{subfigure}
\usepackage{amssymb}
\usepackage{amsmath,bm}
\usepackage{xcolor}
\usepackage{float}
\usepackage{subfigure}
\usepackage[most]{tcolorbox}
\usepackage{soul}
\usepackage{color}
\usepackage[colorlinks=true,linkcolor=blue,allcolors=blue]{hyperref}%
\usepackage[framemethod=TikZ]{mdframed}
\usepackage{lipsum}
\mdfdefinestyle{MyFrame}{%
	linecolor=blue,
	outerlinewidth=2pt,
	roundcorner=20pt,
	innertopmargin=\baselineskip,
	innerbottommargin=\baselineskip,
	innerrightmargin=20pt,
	innerleftmargin=20pt,
	backgroundcolor=gray!50!white}

\begin{document}
	
	
\headsep = 40pt
\title{Pure and Linear Frequency Converter Temporal Metasurface}
\author{Sajjad Taravati and George V. Eleftheriades}
\affiliation{The Edward S. Rogers Sr. Department of Electrical and Computer Engineering, University of Toronto, Toronto, Ontario M5S 3H7, Canada\\
Email: sajjad.taravati@utoronto.ca}
	
\begin{abstract}
Metasurfaces are ultrathin structures which are constituted by an array of subwavelength scatterers with designable scattering responses. They have opened up unprecedented exciting opportunities for extraordinary wave engineering processes. On the other hand, frequency converters have drawn wide attention due to their vital applications in telecommunication systems, health care devices, radio astronomy, military radars and biological sensing systems. Here, we show that a spurious-free and linear frequency converter metasurface can be realized by leveraging unique properties of engineered transmissive temporal supercells. Such a metasurface is formed by time-modulated supercells; themselves are composed of temporal and static patch resonators and phase shifters. This represents the first frequency converter metasurface possessing large frequency conversion ratio with controllable frequency bands and transmission magnitude. In contrast to conventional nonlinear mixers, the proposed temporal frequency converter offers a linear response. In addition, by taking advantage of the proposed surface-interconnector-phaser-surface (SIPS) architecture, a spurious-free and linear frequency conversion is achievable, where all undesired mixing products are strongly suppressed. The proposed metasurface may be digitally controlled and programmed through a field programmable gate array. This makes the spurious-free and linear frequency converter metasurface a prominent solution for wireless and satellite telecommunication systems, as well as invisibility cloaks and radars. This study opens a way to realize more complicated and enhanced-efficiency spectrum-changing metasurface.
\end{abstract}
	
\maketitle

\section{Introduction}

Frequency conversion is a vital task in telecommunication systems, where the frequency of the input signal is translated to a greater or smaller value, i.e., up-converted in transmitters and down-converted in receivers. Practical frequency conversion is desirable to produce large frequency conversion ratios, where the frequency of the input signal is translated from a frequency band to another frequency band, e.g. from UHF-band to L-band. Conventional nonlinear mixers produce unsought spurious signals as a result of the harmonic mixing of the radio-frequency (RF) and local oscillator (LO) signals. For single-tone RF and LO signals, the spurious signal frequencies correspond to the $i \omega_\text{RF}+j \omega_\text{LO}$ harmonic products, with $i$ and $j$ being any integers. However, multi-tone RF and LO signals yield a much more unsought spurious frequencies including the principal $i \omega_\text{RF}+j \omega_\text{LO}$ harmonic products for each RF tone combined with each LO tone individually plus the cross-modulation products between multiple RF and LO tones.

A quasi-pure frequency conversion is conventionally achieved by integration of frequency mixers with bandpass filters which typically result in high insertion loss and large profile structure. Frequency mixers are usually formed by nonlinear components, e.g. Schottky diodes, GaAs FETs and CMOS transistors~\cite{Maas_1993,Henderson_2013,Hashimoto_TMTT_2016}, where the nonlinear response of the component results in generation of an infinite number of mixing products leading to a significant waste of power due to the transition of power to unwanted frequencies. Over the past few decades, several approaches have been proposed to realize harmonic-rejection mixers. Nevertheless, conventional nonlinear mixers, switching mixers, sub-sampling mixers and microwave photonic mixers, even in their most ideal operation regimes, still suffer from unwanted mixing products~\cite{Maas_1993,Naviner_1999,Pekau_2005,Jiang_OE_2017}.

Space-time refractive-index modulation represents a prominent alternative approach for the realization of linear frequency mixers. Lately, space-time-modulated media have attracted a surge of interest thanks to their extraordinary capability in multifunctional operations, e.g., mixer-duplexer-antennas~\cite{Taravati_LWA_2017}, unidirectional beam splitters~\cite{Taravati_Kishk_PRB_2018}, nonreciprocal filters~\cite{wu2019isolating}, and signal coding metagratings~\cite{taravati_PRApp_2019}. In addition, a large number of versatile and high efficiency electromagnetic systems have been recently reported based on the unique properties of space-time modulation, including space-time metasurfaces for advanced wave engineering and extraordinary control over electromagnetic waves~\cite{Fan_APL_2016,Fan_mats_2017,Salary_2018,Taravati_Kishk_TAP_2019,zang2019nonreciprocal_metas,inampudi2019rigorous,elnaggar2019generalized,wang2018photonic,Grbic2019serrodyne,Taravati_Kishk_MicMag_2019,ptitcyn2019time,du2019simulation,wang2019multifunctional,taravati2019full,li2020time,taravati_STMetasRev_2020}, nonreciprocal platforms~\cite{wentz1966nonreciprocal,Taravati_PRB_2017,Taravati_PRB_SB_2017,Taravati_PRAp_2018,oudich2019space,chegnizadeh2020non}, frequency converters~\cite{Taravati_PRB_Mixer_2018,Grbic2019serrodyne}, and time-modulated antennas~\cite{shanks1961new,zang2019nonreciprocal}. Such strong capabilities of space-time-modulated media is due to their unique interactions with the incident field~\cite{Taravati_PRB_2017,li2019nonreciprocal,correas2018magnetic,liu2018huygens,du2019simulation,elnaggar2019modelling,Taravati_AMA_PRApp_2020}.

\begin{figure*}
	\begin{center}
		\includegraphics[width=1.5\columnwidth]{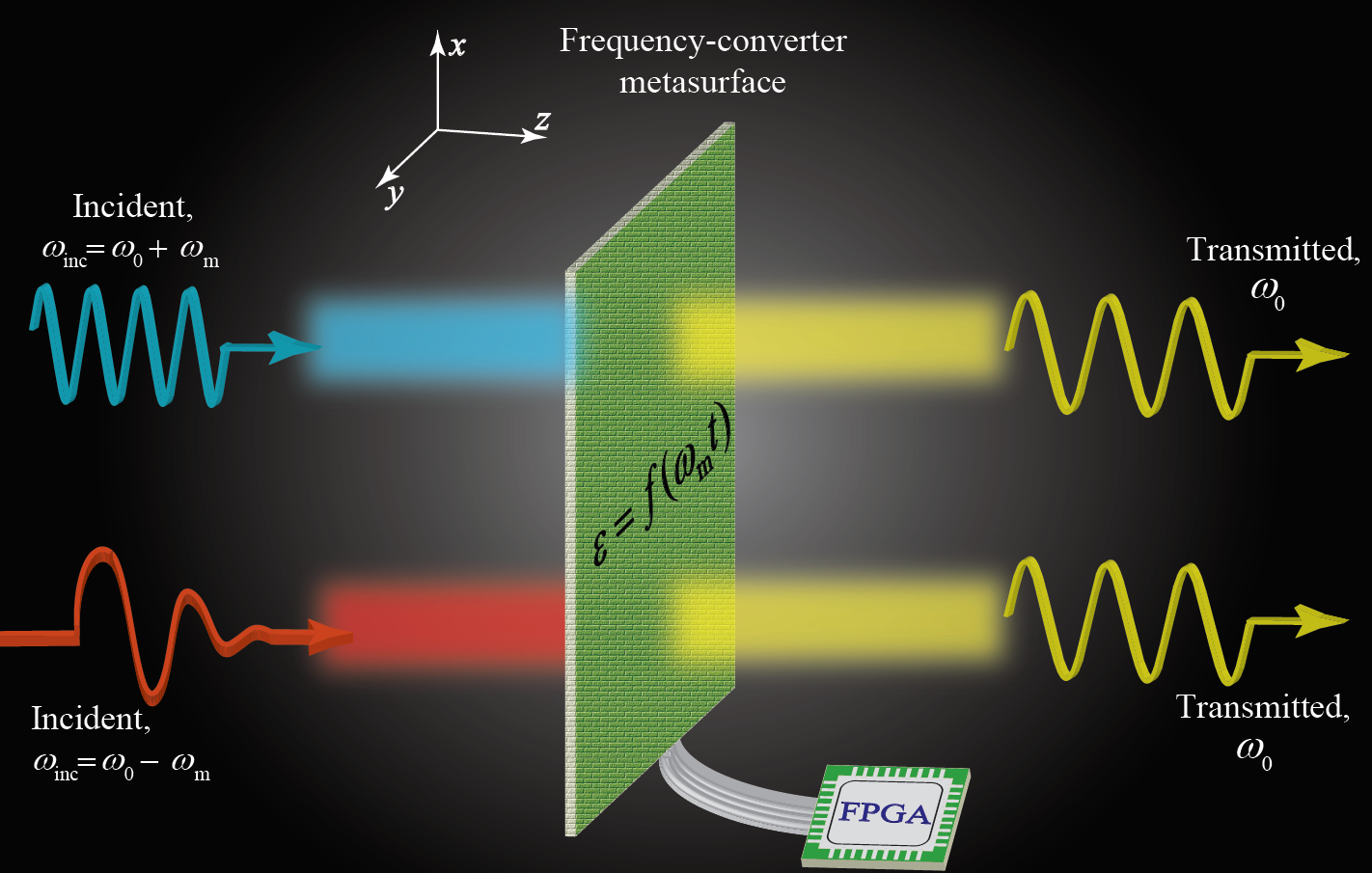}
		\caption{The conceptual illustration of programmable spurious-free and linear frequency conversion induced by a temporal metasurface.}
		\label{fig:sch} 
	\end{center}
\end{figure*}

Periodic time modulation is a promising approach for \textit{linear} frequency mixing thanks to the lack of harmonics of the incident frequency. This is due to the fact that the temporal medium is only periodic with respect to the modulation frequency $\omega_\text{m}$. Time modulation provides the required energy for transition from the fundamental temporal frequency $\omega_0$ to an infinite number of time frequency harmonics $\omega_\text{O}=\omega_0 \pm n\omega_\text{m}$, and hence, the output wave includes time harmonics of the modulation wave~\cite{Oliner_PIEEE_1963,Taravati_PRB_2017,Taravati_PRB_Mixer_2018}. Although the lack of harmonics of the incident signal in the output signal represents an advantage of time-modulated mixers over conventional nonlinear mixers, the existence of harmonics of the modulation in the output signal is still problematic and indicates a non-pure frequency conversion and waste of energy~\cite{Grbic2019serrodyne}. Another drawback of previously reported time-modulated mixers is their extra small frequency conversion ratio~\cite{Grbic2019serrodyne}, which is impractical especially for telecommunication systems, where large frequency conversions are required. 

Herein, we introduce a technique for the realization of spurious-free and linear frequency converter metasurfaces based on time modulation technique, where harmonic electromagnetic transitions in temporally-periodic systems are prohibited by \textit{tailored} photonic band gaps introduced by the engineered spatial-aperiodicity of the supercells. The proposed surface-interconnector-phaser-surface (SIPS) architecture acts inherently as a multi-band bandpass filter, i.e., providing multichannel bandpass transmissions and strong stopbands which are leveraged for local suppression of undesired time harmonics. Such local suppressions lead to a spurious-free and linear frequency conversion and inhibits the leakage of incident wave to undesired time harmonics which would have yielded substantial waste of power. The proposed frequency converter metasurface inherits the linearity property of time modulation. In addition, the frequency bands of the frequency converter metasurface may be controlled via time modulation parameters as well as patch resonators.

\section{Results}\label{sec:oper}
Figure~\ref{fig:sch} presents the conceptual illustration of programmable spurious-free and linear frequency conversion induced by a temporal metasurface. The metasurface is characterized with the general time-modulated permittivity of $\epsilon(t)=f(\omega_\text{m} t)$, where $f(.)$ represents a periodic function. The metasurface is illuminated by a plane wave from the left side. We show that with proper design of the metasurface supercells, a spurious-free and linear frequency up-conversion, from $\omega_0-\omega_\text{m}$ to $\omega_0$, and a spurious-free and linear frequency down-conversion, from $\omega_0+\omega_\text{m}$ to $\omega_0$ can be achieved. The functionality and frequencies of the metasurface can be controlled via a field-programmable gate array (FPGA).

\begin{figure*}
	\begin{center}
		\subfigure[]{\label{fig:p1} 
			\includegraphics[width=1.3\columnwidth]{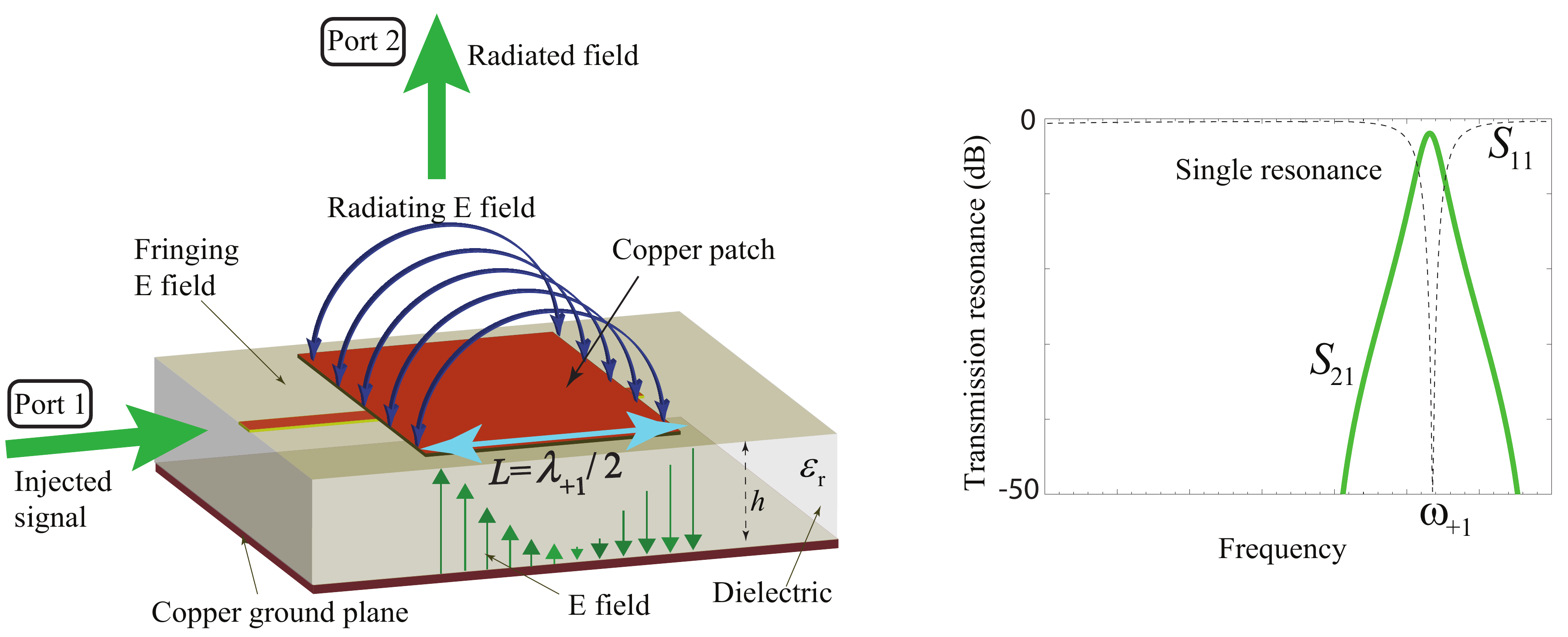}} 
		\subfigure[]{\label{fig:p2} 
			\includegraphics[width=1.7\columnwidth]{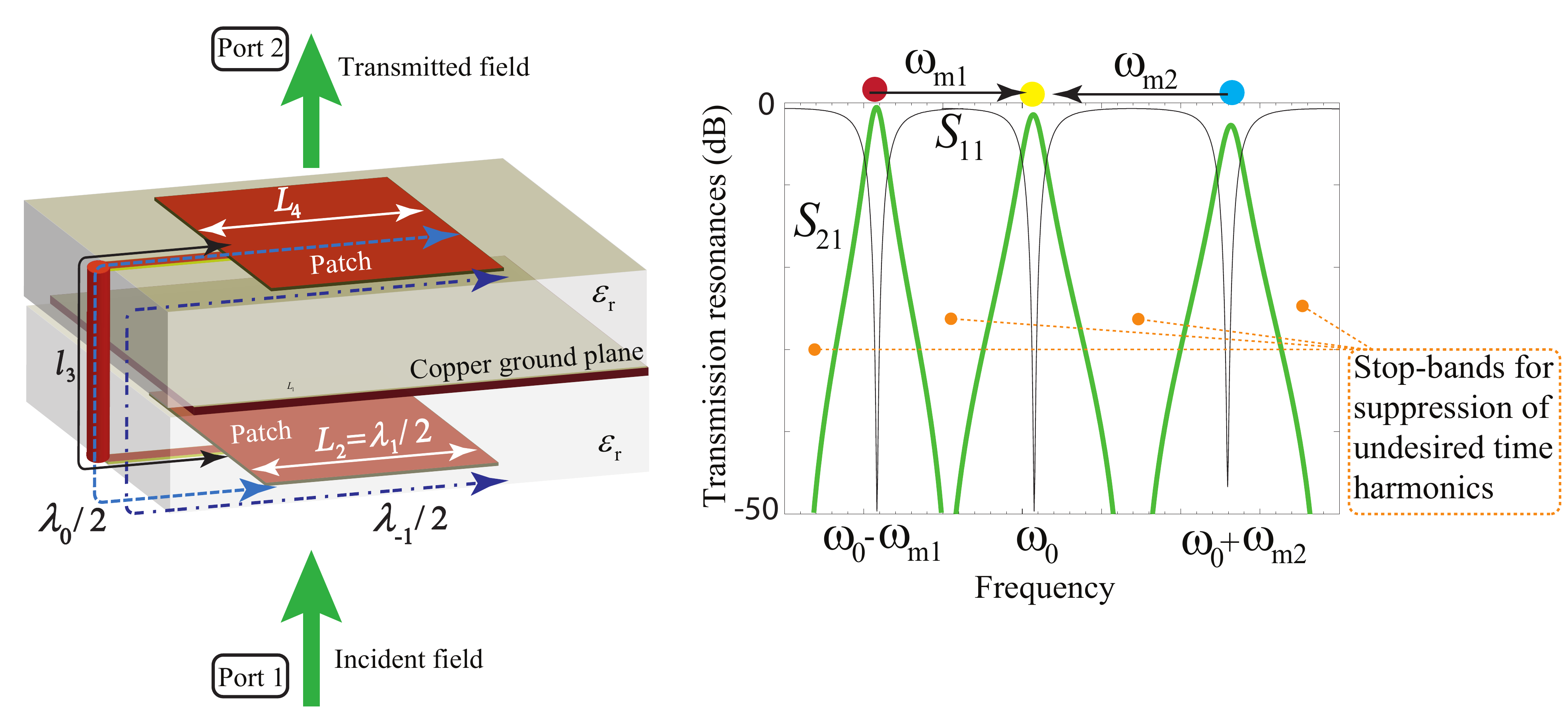}}
		\caption{Transmission resonances in (a) a standard patch antenna introducing single transmission resonance. (b) Surface-interconnector-phaser-surface (SIPS) architecture composed of two interconnected microstrip patch resonators, introducing multiple in-band full-transmission resonances as well as strong out-of-band reflection.} 
		\label{Fig:pp}
	\end{center}
\end{figure*}

Figure~\ref{fig:p1} shows wave propagation and radiation in a standard half-wavelength microstrip patch resonator antenna where the injected signal in the feed line of the patch antenna is efficiently radiated to air. Here, the patch resonator antenna introduces a single resonance providing full-transmission from the feeding line to air at $\omega_{1}$ corresponding to the resonant frequency of the patch antenna. At this resonance, the length of the patch antenna reads half-wavelength of the incident frequency, i.e. $L=\lambda_{1}/2$.

Figure~\ref{fig:p2} depicts a non-modulated single-fed surface-interconnection-phaser-surface (SIPS) supercell where two single-fed patch resonators are interconnected in a three-layer architecture, composed of two conductor layers mounted on two dielectric layers and shielded from each other by a copper ground plane layer. In contrast to the single patch resonator in Fig.~\ref{fig:p1}, here the structure introduces at least three major resonances, corresponding to the single patch resonance and coupled structure resonances, as shown in Fig.~\ref{fig:p2}. As a result, the incident field from the bottom (top) of the structure is transmitted to the top (bottom) of the SIPS architecture at three different frequencies, i.e., $\omega_{-1}=\omega_{0}-\omega_\text{m1}$, $\omega_0$ and $\omega_{+1}=\omega_{0}+\omega_\text{m2}$. Here, $\omega_\text{m1}$ represents the local frequency for down-conversion and $\omega_\text{m2}$ represents the local frequency for up-conversion. As a result, with proper design of the SIPS structure, controllable full-transmission passbands can be achieved at the desired frequencies, and stopbands exhibiting large suppressions can be achieved at undesired frequencies, e.g., $\omega_{-2}$, $\omega_{-3}$, $\omega_{+2}$, $\omega_{+3}$, etc. This property of the proposed SIPS architecture offers an outstanding opportunity for spurious-free and linear frequency conversion when integrated with time modulation.

\begin{figure}
	\begin{center}
		\subfigure[]{\label{Fig:unf_p} 
			\includegraphics[width=1\columnwidth]{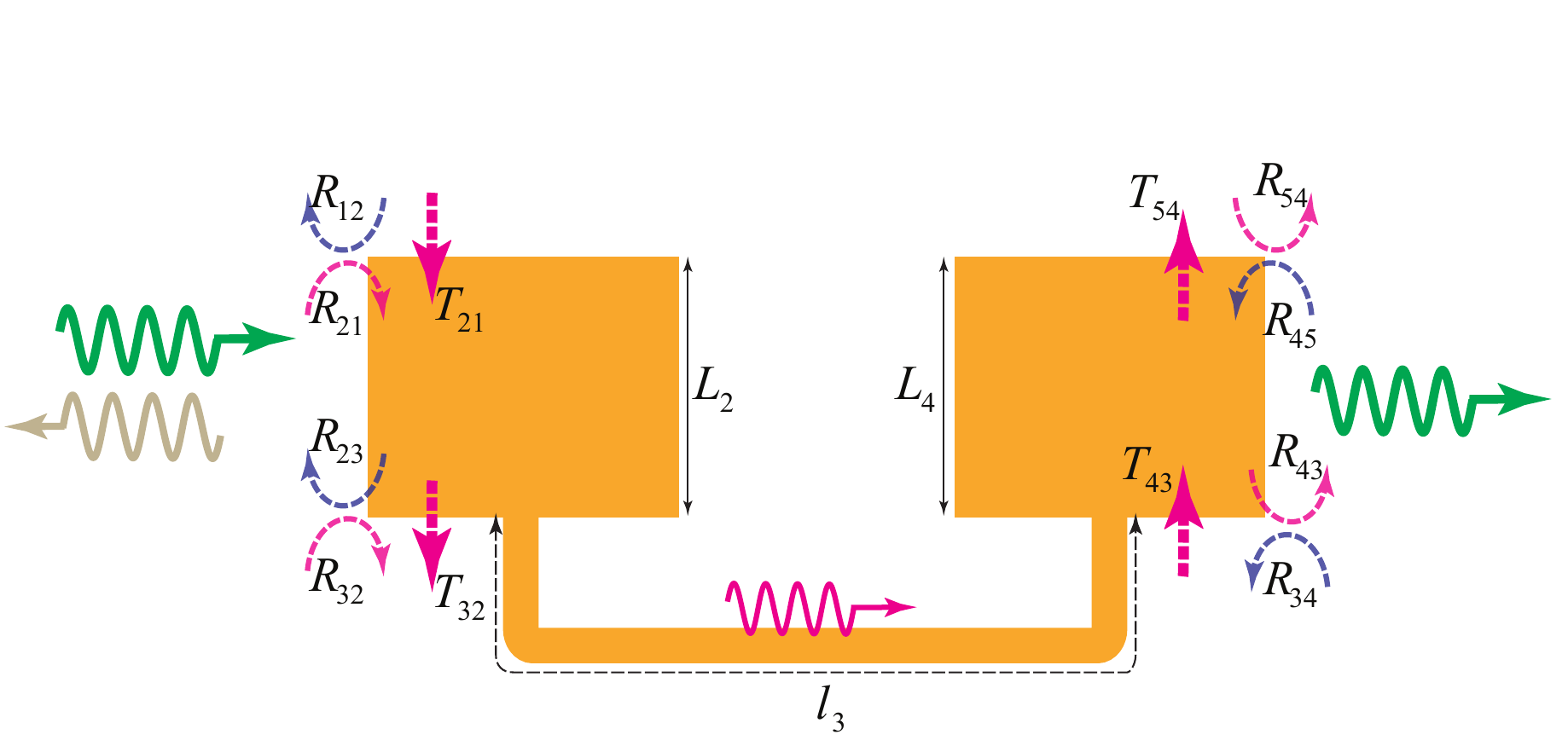}}
		\subfigure[]{\label{Fig:unf_p_b}
			\includegraphics[width=1\columnwidth]{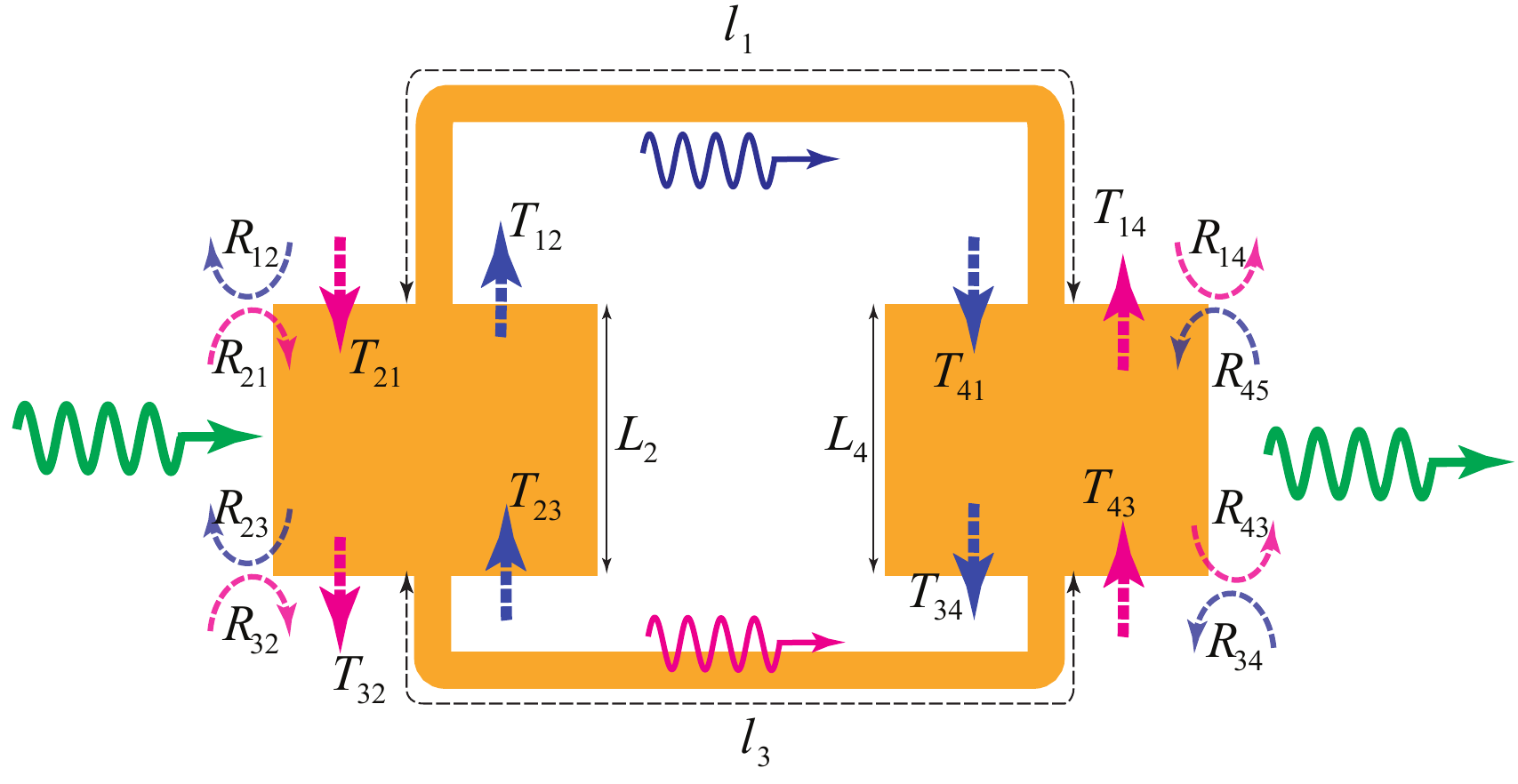}}
		\caption{Total transmission following multiple transmissions and reflections inside (a) Single-interconnected single-fed patch antenna pair. (b) Double-interconnected double-fed patch antenna pair.} 
		\label{Fig:unf}
	\end{center}
\end{figure}

Figure~\ref{Fig:unf_p} shows an unfolded version of the single-fed SIPS structure in Fig.~\ref{fig:p2}, where a pair of single-fed patch resonators with lengths $L_2$ and $L_4$ are interconnected via interconnections of length $l_3$. The wavenumbers $k$ of such inhomogeneous microstrip transmission lines depend on their width~\cite{Garg_2001}. Hence, the wavenumbers are different in each region, that is, $k_1(=k_5)\neq k_2(=k_4) \neq k_3$. The electric field in the $m$th region ($m=1,\ldots,5$) is composed of forward and backward waves as $E_m=V_m^+ e^{-j k_m z}+ V_m^- e^{j k_m z}$, where $V_m^+$ and $V_m^-$ are the amplitudes of the forward and backward waves, respectively, and $k_m$ is the wavenumber. It should be noted that the backward waves, propagating along the $-z$ direction, are due to reflection at the different interfaces between adjacent regions. Upon application of boundary conditions at the interface between regions $m$ and $m+1$, the total transmission and total reflection coefficients between regions $m$ and $m+1$ are found as as~\cite{Chew_1995}
\begin{subequations}
	\begin{equation}
	\widetilde{T}_{m+1,m} = \frac{V_{m+1}^+}{V_{m}^+}=\frac{T_{m+1,m} e^{-j (k_{m}-k_{m+1}) z}}{1-R_{m+1,m} \widetilde{R}_{m+1,m+2} e^{-j 2 k_{m+1} l_{m+1}}},
	\label{eqa:xxx}
	\end{equation}
	\begin{equation}
	\widetilde{R}_{m,m+1} = \frac{R_{m,m+1}+\widetilde{R}_{m+1,m+2} e^{-j 2k_{m+1} l_{m+1}}}{1+R_{m,m+1} \widetilde{R}_{m+1,m+2} e^{-j 2 k_{m+1} l_{m+1}}}.
	\label{eqa:www}
	\end{equation}
	\label{eqa:TRtotal}
\end{subequations}
where $R_{m,m+1}=(\eta_{m+1}-\eta_{m})/(\eta_{m+1}+\eta_{m})$, with $\eta_m$ being the intrinsic impedance of region~$m$, is the local reflection coefficient within region $m$ between regions $m$ and $m+1$, and $R_{m+1,m}=-R_{m,m+1}$. The local transmission coefficient from region $m$ to region $m+1$ reads $T_{m+1,m}=1+R_{m,m+1}$. It should be noted that the term $e^{-j (k_{m}-k_{m+1}) z}$ in~\eqref{eqa:xxx} indicates that, due to the nonuniformity of the structure in Fig.~\ref{fig:p1}, a phase shift occurs at each interface which corresponds to the difference between the wavenumbers in adjacent regions. We assume ${k_1=k_5=k_0}$, ${k_2=k_4=k_\text{p}}$ and ${k_3=k_\text{t}}$ the wavenumbers in the air, in the two patches, and in the interconnecting transmission line, respectively. We consider $L_2=L_4$, $R_{1,2}=-R_{2,1}=-R_{4,5}=R_\text{p}$, and $R_{2,3}=-R_{3,2}=-R_{3,4}=R_{4,3}=R_\text{t}=(\eta_\text{t}-\eta_\text{p})/(\eta_\text{t}+\eta_\text{p})$ the local reflection coefficient at the interface between a patch and the interconnecting transmission line.

Then, the total transmission coefficient for the single-fed SIPS metasurface of Fig.~\ref{Fig:unf_p}, from region 1 to region 5, reads 

\begin{equation}
S_\text{21}=T_\text{5,1} = \prod_{m=1}^{4} \widetilde{T}_{m+1,m} e^{-j k_m l_m},
\label{eqa:mm2}
\end{equation}

\noindent where $\widetilde{T}_{m+1,m}$, for $m=1,\ldots,4$ is provided by~\eqref{eqa:xxx} with~\eqref{eqa:www}. The total transmission coefficient from the non-modulated single-fed SIPS supercell in Fig.~\ref{Fig:unf_p} is found in terms of local reflection coefficients as
\begin{equation}
S_\text{21}= \frac{(1-R_\text{12}^2) (1-R_\text{23}^2) e^{j (k_\text{p}+ k_\text{0}-2k_\text{3}) l_3 }  }{ (R_\text{12}R_\text{23}+e^{j 2k_\text{p} L } )^2 -  (R_\text{23} e^{j 2k_\text{p} L}+R_\text{12} )^2 e^{-j 2k_\text{3} l_3}}.
\label{eqa:S21_a}
\end{equation}

\noindent The term $e^{-j 2k_\text{t} l_3}$ in the denominator of this expression corresponds to the round-trip propagation through the middle transmission line, whose multiplication by $e^{j 4k_\text{p} L }$ in the adjacent bracket corresponds to the patch-line-patch coupled-structure resonance, with length $2L+l_3$.

Figure~\ref{Fig:unf_p_b} shows the unfolded version of the double-fed SIPS structure, where a pair of double-fed patch resonators with lengths $L2$ and $L_4$ are double-interconnected via two interconnections possessing different lengths, i.e., $l_1$ and $l_2$. We assume that the upper interconnection with length $l_2$ represents the region 1. The total transmission coefficient from the double-fed SIPS structure in Fig.~\ref{Fig:unf_p_b} reads
\begin{equation}
\begin{split}
&S_\text{21}= \frac{(1-R_\text{12}^2) (1-R_\text{23}^2) e^{j (k_\text{p}+ k_\text{0}-2k_\text{3}) l_3 }  }{ (R_\text{12}R_\text{23}+e^{j 2k_\text{p} L } )^2 -  (R_\text{23} e^{j 2k_\text{p} L}+R_\text{12} )^2 e^{-j 2k_\text{3} l_3}}\\
&\quad-\frac{(1-R_\text{32}^2) (1-R_\text{21}^2) e^{j (k_\text{p}+ k_\text{0}-2k_\text{1}) l_1 }  }{ (R_\text{32}R_\text{21}+e^{j 2k_\text{p} L } )^2 -  (R_\text{21} e^{j 2k_\text{p} L}+R_\text{32} )^2 e^{-j 2k_\text{1} l_1}}.
\label{eqa:S21_b}
\end{split}
\end{equation}

Equation~\eqref{eqa:S21_b} highlights the effect of $l_1$ and $l_3$ on the transmission through the double-fed SIPS in Fig.Figure~\ref{Fig:unf_p_b}. Hence, by changing the phase shift through $l_1$ and $l_3$ the frequency and magnitude of the transmission parameter $S_{21}$ is controlled. In general, fixed transmission-line-based phase shifters with lengths $l_1$ and $l_3$ can be replaced by two digitally controlled phase shifters.

\begin{figure}
	\begin{center}
		\includegraphics[width=1\columnwidth]{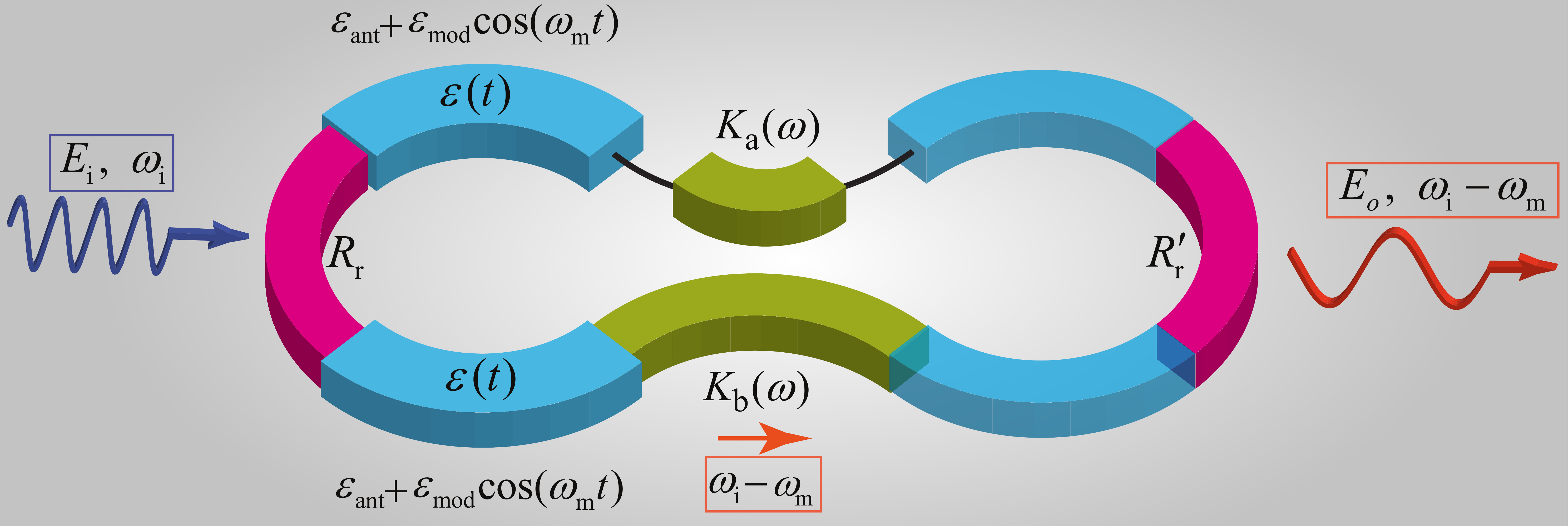}
		\caption{Schematic representation and operation principle of the time-modulated SIPS radiating supercell.}
		\label{fig:coupl} 
	\end{center}
\end{figure}

Figure~\ref{fig:coupl} sketches the wave propagation and transmission through the time-modulated supercell, characterized with two time-varying resonators possessing a periodic time-dependent permittivity, i.e., 
\begin{equation}\label{eqa:perm}
\epsilon(t)=\epsilon_\text{ant} + \epsilon_\text{mod} \cos(\omega_\text{m} t),
\end{equation}
where $\epsilon_\text{ant}$ is the average effective permittivity of the patch antennas, $\epsilon_\text{mod}$ denotes the modulation amplitude and $\omega_\text{m}$ denotes the modulation frequency. The electric field inside each of these two time-varying resonators may be expressed based on the superposition of two supported space-time harmonic fields, i.e.,
\begin{equation}\label{eqa:el}
E_\text{m}(\zeta,t)=a(\zeta) e^{-i \left(k_\text{a} \zeta -\omega_\text{0} t \right)}+b(\zeta) e^{-i \left(k_\text{b} \zeta -(\omega_\text{0}+\omega_\text{m}) t \right)}.
\end{equation}

\begin{figure}
	\begin{center}
		\subfigure[]{\label{fig:circ1} 
			\includegraphics[width=1\columnwidth]{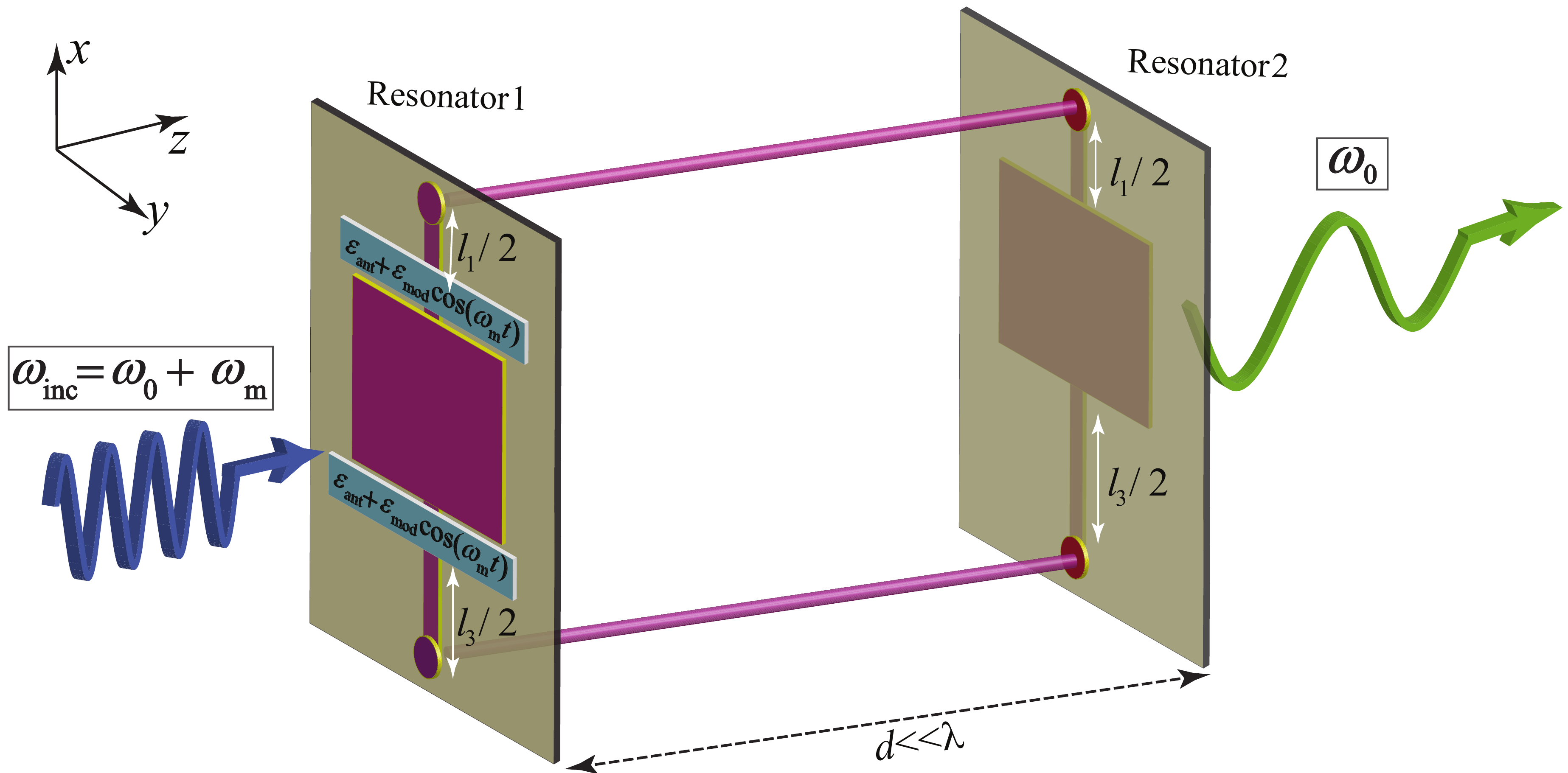}}
		\subfigure[]{\label{fig:circ2} 
			\includegraphics[width=1\columnwidth]{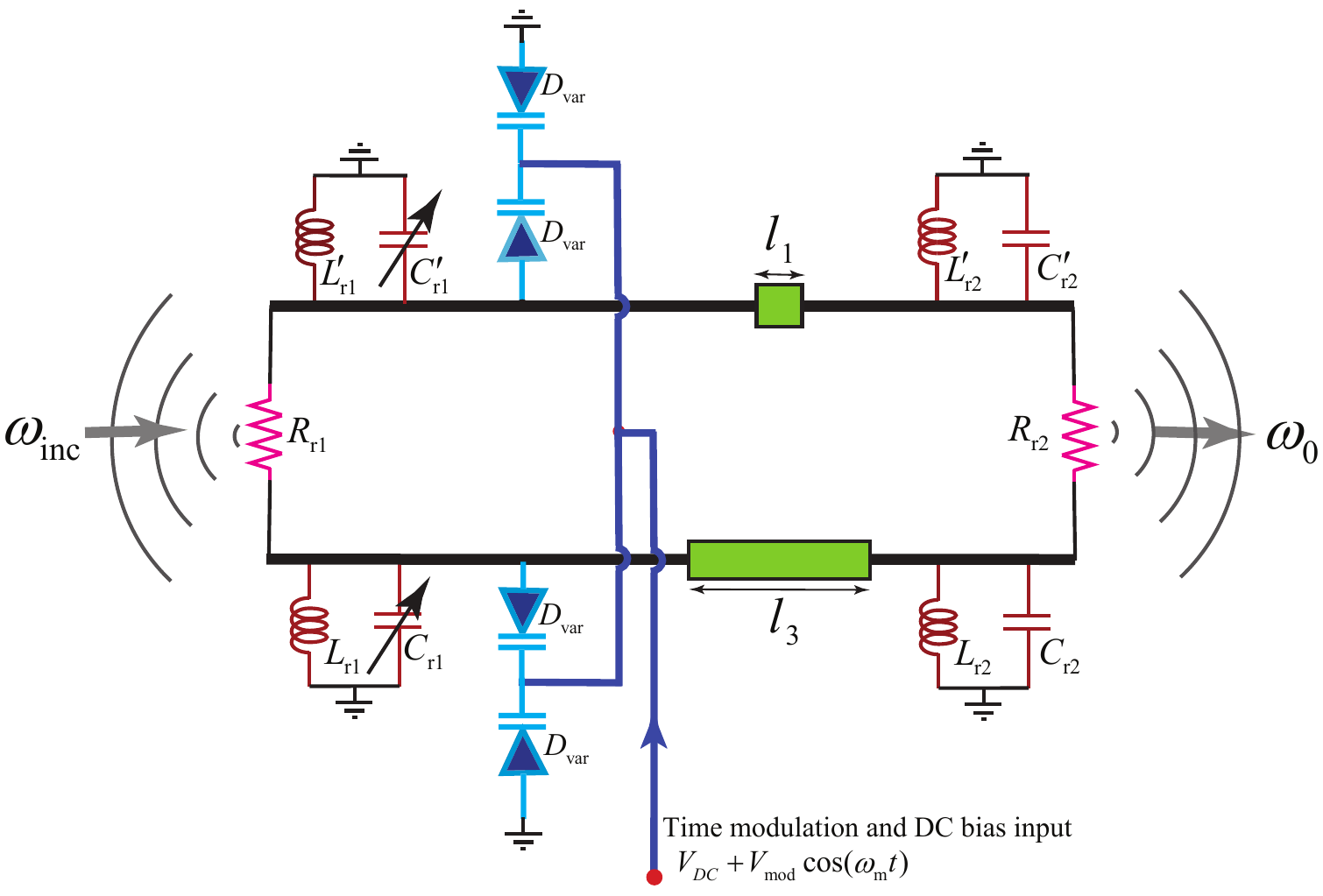}}	
		\caption{Time-modulated SIPS supercell. (a) Realization using a temporal double-fed microstrip patch resonator and a non-modulated double-fed microstrip patch resonator. (b) Circuit model and realization of the temporal modulation using RF-biased varactors.}
		\label{fig:circ} 
	\end{center}
\end{figure}

The corresponding wave equation reads $c^2 \partial^{2} \textbf{E}/ \partial \zeta^{2}= \partial^{2} [\epsilon_\text{eq}(t) \textbf{E}]/\partial t^{2}$. Inserting the electric field in~\eqref{eqa:el} into the wave equation results in 
\begin{equation}
\begin{split}
&\left( \frac{\partial^{2} }{\partial \zeta^{2}}  \right) \left[  a(\zeta) e^{-i \left(k_\text{a} \zeta -\omega_\text{0} t \right)}+b(\zeta) e^{-i \left( k_\text{b}\zeta -(\omega_\text{0}+\omega_\text{m}) t \right)}   \right]\\
& \qquad  = \frac{1}{c^2} \frac{\partial^{2} }{\partial t^{2}} 
\left(\bigg[\epsilon_\text{ant} + \frac{\delta}{2} e^{i(\omega_\text{m} t)}  + \frac{\delta}{2} e^{-i(\omega_\text{m} t)} \right]  \\
&\qquad \quad \left(a(\zeta) e^{-i \left(k_\text{a} \zeta -\omega_\text{0} t \right)}+b(\zeta) e^{-i \left( k_\text{b}\zeta -(\omega_\text{0}+\omega_\text{m}) t \right)} \right) \bigg),
\end{split}
\end{equation}
and applying the space and time derivatives, while using a slowly varying
envelope approximation, multiply both sides with $\exp({i \left[k_\text{a} \zeta -\omega_\text{0} t \right]})$, which gives
\begin{equation}\label{eqa:eq2}
\begin{split}
& \left[k_\text{a}^2 a(\zeta) -2ik_\text{a} \frac{d a(\zeta)}{d \zeta} \right] +\left[ k_\text{b}^2  b(\zeta) -2i(k_\text{b}) \frac{d b(\zeta)}{d \zeta} \right] e^{i \omega_\text{m} t }\\
& = \frac{1}{c^2} 
\bigg(\bigg[\omega_\text{0}^2\epsilon_\text{ant}+ (\omega_\text{0}+\omega_\text{m})^2 \frac{\delta}{2} e^{i(\omega_\text{m} t)}  \\
& \qquad \qquad \qquad \qquad \qquad \qquad+ (\omega_\text{0}-\omega_\text{m})^2\frac{\delta}{2} e^{-i(\omega_\text{m} t)} \bigg] a(\zeta)\\
&+\bigg[\omega_\text{0}^2 \epsilon_\text{ant}
+ (\omega_\text{0}+2\omega_\text{m})^2\frac{\delta}{2}  e^{i(\omega_\text{m} t)} + \omega_\text{0}^2 \frac{\delta}{2}  e^{-i(\omega_\text{m} t)} \bigg] b(\zeta) e^{i \omega_\text{m} t }  \bigg),
\end{split}
\end{equation}
Then, we apply $\int_{0}^{\frac{2\pi}{\omega_\text{m}}}dt$ to both sides of~\eqref{eqa:eq2}, which yields a coupled differential equation for the field coefficients, i.e., 
\begin{equation}\label{eqa:coupl}
	\frac{d }{d \zeta}  \begin{bmatrix} a(\zeta) \\ b(\zeta) \end{bmatrix}
	=
	\begin{bmatrix}
	M_0 & C_0 \\ C_{1}& M_{1}
	\end{bmatrix}
	\begin{bmatrix} a(\zeta) \\ b(\zeta) \end{bmatrix},
\end{equation}
where $M_0=i (k_\text{ant}^2-k_\text{a}^2 )/(2 k_\text{a})$, $M_{1}=i(k_0^2-k_\text{b}^2)/(2 k_\text{b})$, $C_0=i\delta k_0^2/(4 k_\text{a})$, $C_{1}=i\delta k_0^2/(4 k_\text{b})$. The solution to the coupled differential equation in~\eqref{eqa:coupl} reads
\begin{subequations}\label{eqa:coup_sol}
	\begin{align}
	a(\zeta)=\frac{E_0}{2 \varDelta}&\bigg((M_0-M_{1}+\varDelta) e^{\frac{M_0+M_{1}+\varDelta}{2}\zeta} \\
	&- (M_0-M_{1}-\varDelta) e^{\frac{M_0+M_{1}-\varDelta}{2}\zeta}  \bigg),
	\end{align}
	\begin{align}
	b(\zeta)=\frac{E_0 C_{1}}{ \varDelta} \left( e^{\frac{M_0+M_{1}+\varDelta}{2}\zeta}  - e^{\frac{M_0+M_{1}-\varDelta}{2}\zeta}  \right),
	\end{align}
\end{subequations}
where $\varDelta=\sqrt{(M_0-M_{1})^{2}+4 C_0C_{1}}$.

\begin{figure*}
	\begin{center}
		\subfigure[]{\label{fig:arch} 
			\includegraphics[width=1.6\columnwidth]{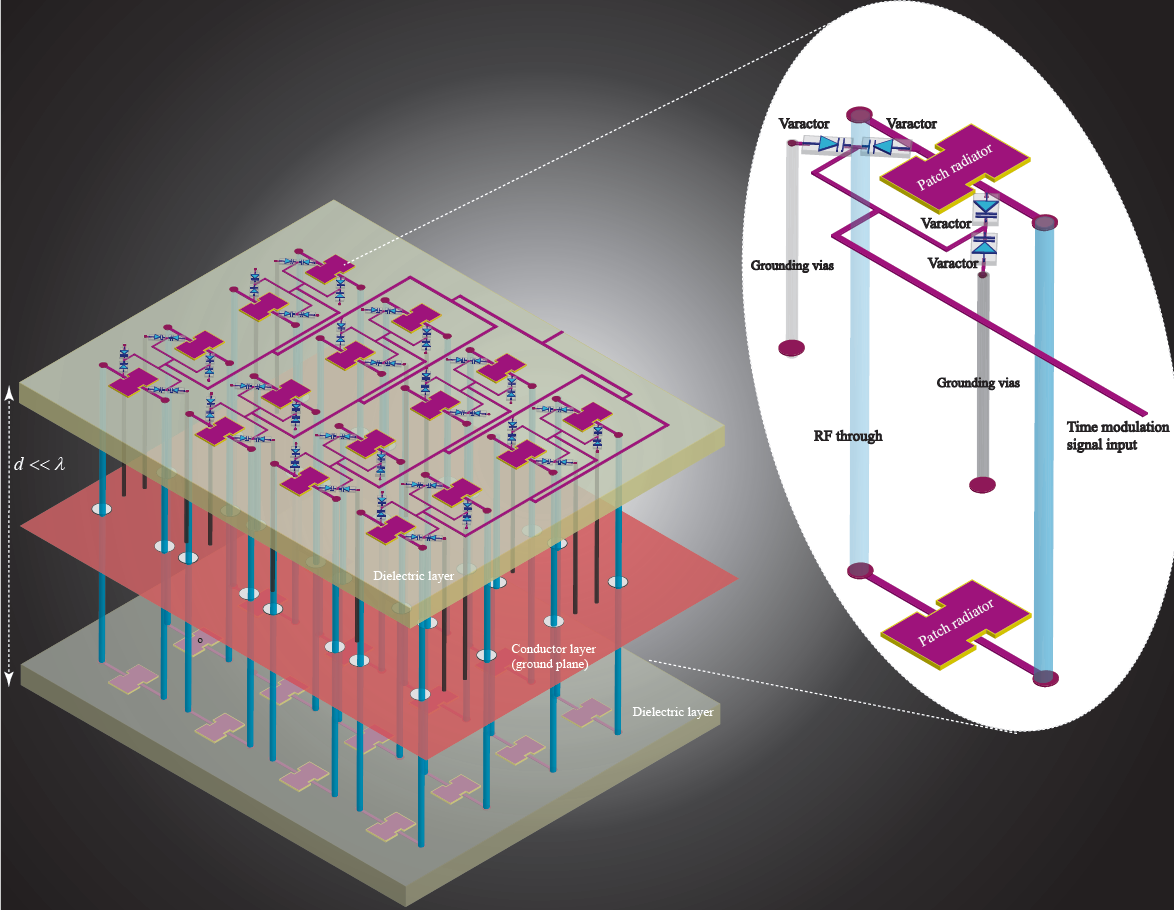}}
		\subfigure[]{\label{Fig:ph} 
			\includegraphics[width=1.5\columnwidth]{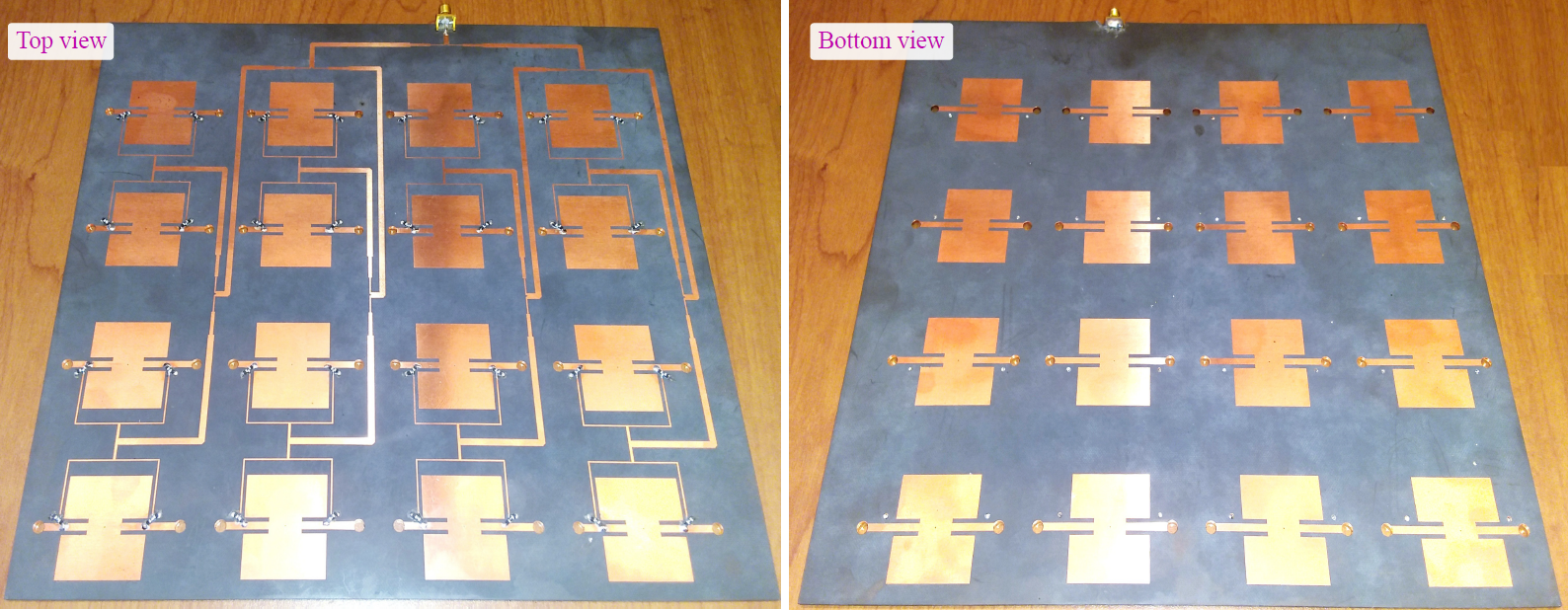}}
		\caption{Frequency converter temporal metasurface. (a)~Architecture of the complete structure. (b)~Photos of the fabricated metasurface. (left)~Top view. (right)~Bottom view.} 
		\label{Fig:arcph}
	\end{center}
\end{figure*}

Figure~\ref{fig:circ1} depicts the realization of the time-modulated SIPS supercell using a temporal double-fed microstrip patch resonator and a non-modulated double-fed microstrip patch resonator. Figure~\ref{fig:circ2} shows a circuit model and realization of the temporal modulation using RF-biased varactors. Figure~\ref{fig:arch} illustrates the architecture of the fabricated proof-of-principle frequency converter temporal metasurface. Figure~\ref{Fig:ph} provides two photos showing the top and bottom views of the fabricated metasurfaces. The metasurface is realized using multilayer circuit technology, where two $10$~in $\times$~$10.5$~in RT5880 substrates with thickness $h = 31$~mil are assembled to realize a three metalization layer structure. The permittivity of the substrate is $\epsilon=\epsilon_\text{r}(1-j \tan \delta)$, with $\epsilon_\text{r} = 2.2$ and $\tan \delta = 0.0009$ at 10 GHz. The middle conductor of the structure (shown in Fig.~\ref{fig:arch}(a)) acts as the RF ground plane for the patch antennas and transmission lines. The DC bias and the modulation signal are both delivered to the supercells through the top conductor layer. One may deliver the DC bias and the modulation signal through the middle conductor sheet. Each side of the metasurface includes 16 microstrip patch resonators, where the dimensions of the 2$\times$16 microstrip patches are $1.18$~in $\times$~$1.417$~in. The connections between the conductor layers are provided by an array of circular metalized through holes, where 32 vias of 40~mil diameter provide grounding point at the top and bottom layers for varactors. Additionally, 16 vias of 20~mil diameter are placed exactly at the center of patch resonator ensuring a DC null on the patches so that proper reverse-bias operation of varactors are guaranteed. Furthermore, the RF path connection between the two sides of the metasurface is provided by 32 via holes, with optimized dimensions of $157$~mils for the via diameters, and $320$~mils for the hole diameter in the via middle conductor ensuring that the RF through from the top layer to the bottom layer is safely isolated from the middle ground-plane conductor. For the varactors, we have utilized 64 number of BB837 silicon tuning varactor diodes manufactured by the Infineon Technologies.

\begin{figure}
	\begin{center}
		\subfigure[]{\label{fig:meas1} 
			\includegraphics[width=1\columnwidth]{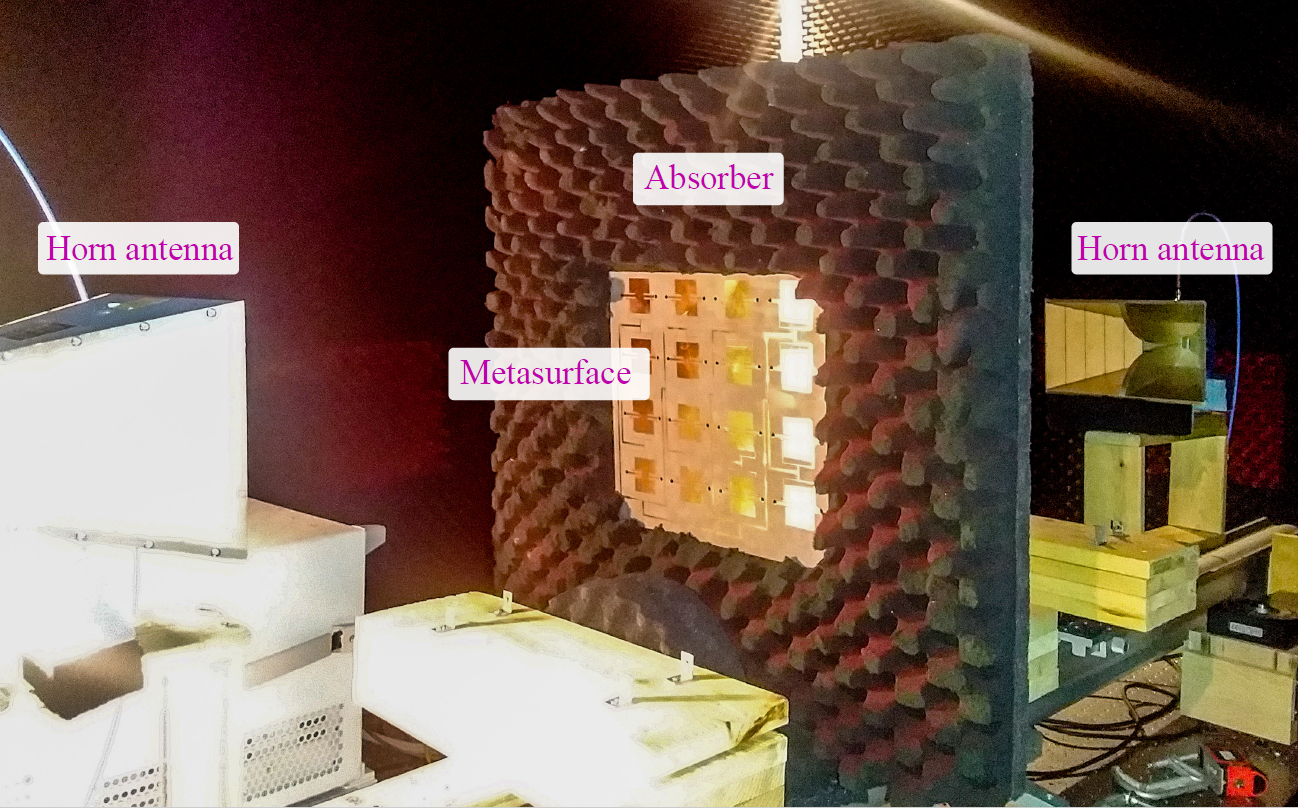}} 
		\subfigure[]{\label{fig:meas2} 
			\includegraphics[width=1\columnwidth]{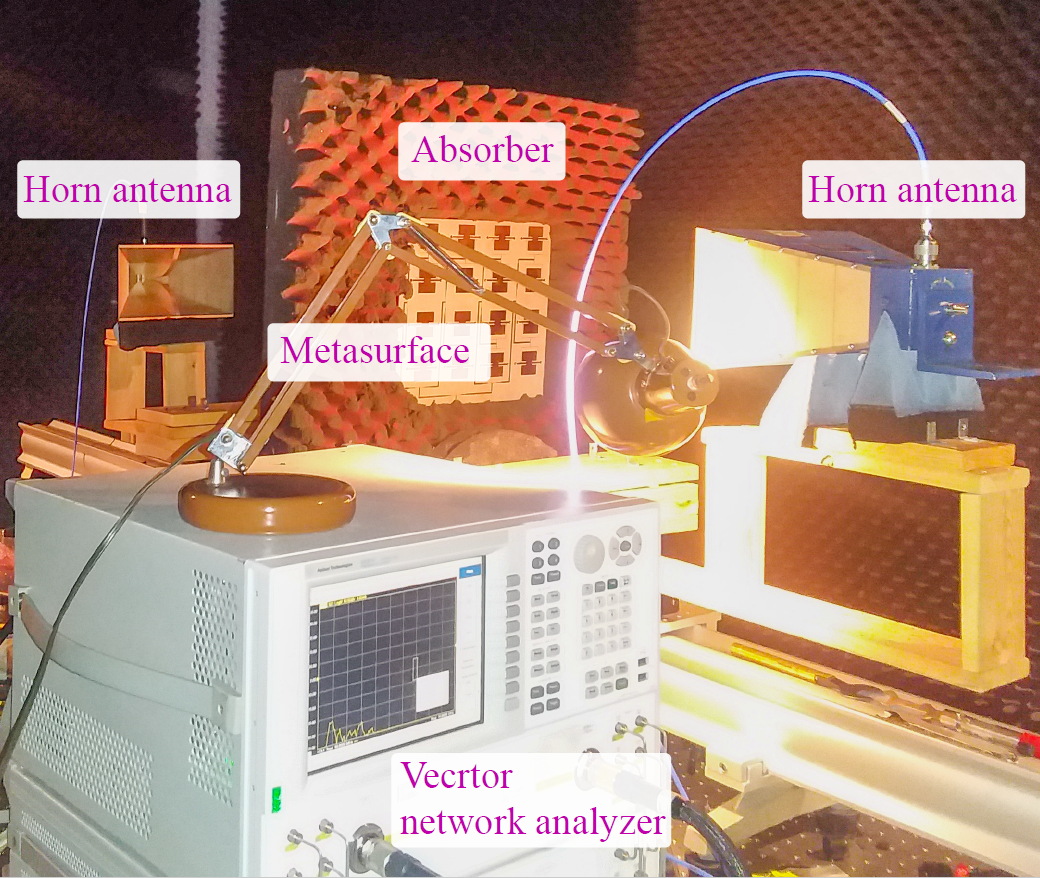}}
		\caption{Experimental set-up for measuring the scattering parameters of the static metasurface, i.e., $\omega_\text{m}=0$.} 
		\label{Fig:vna}
	\end{center}
\end{figure}

Figures~\ref{fig:meas1} and~\ref{fig:meas2} show the experimental set-up for measuring the scattering parameters of the static metasurface, i.e., $\omega_\text{m}=0$.

\begin{figure}
	\begin{center}
		\subfigure[]{\label{fig:s21} 
			\includegraphics[width=0.7\columnwidth]{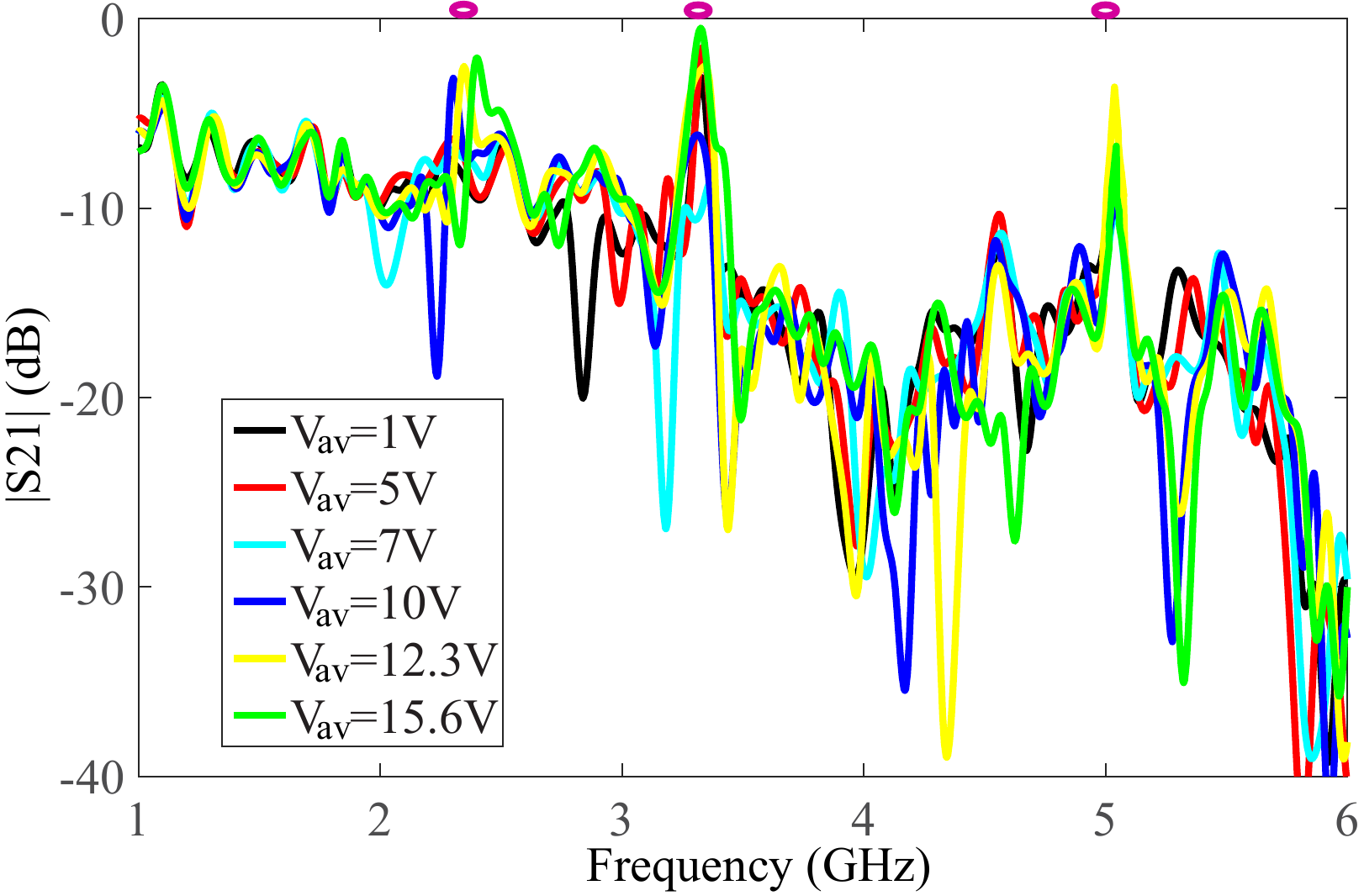}} 
		\subfigure[]{\label{fig:s11} 
			\includegraphics[width=0.7\columnwidth]{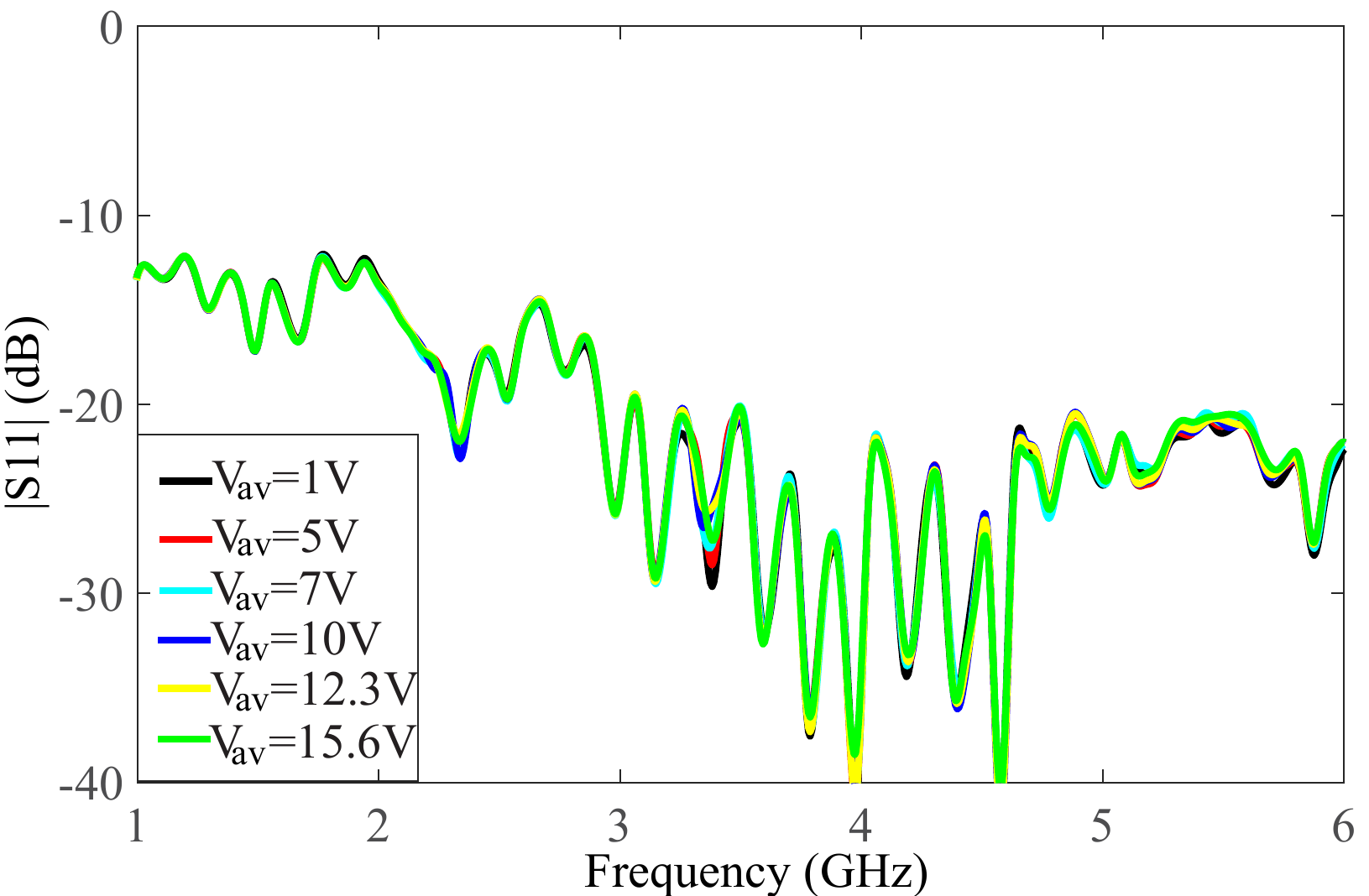}}
		\subfigure[]{\label{fig:s22} 
			\includegraphics[width=0.7\columnwidth]{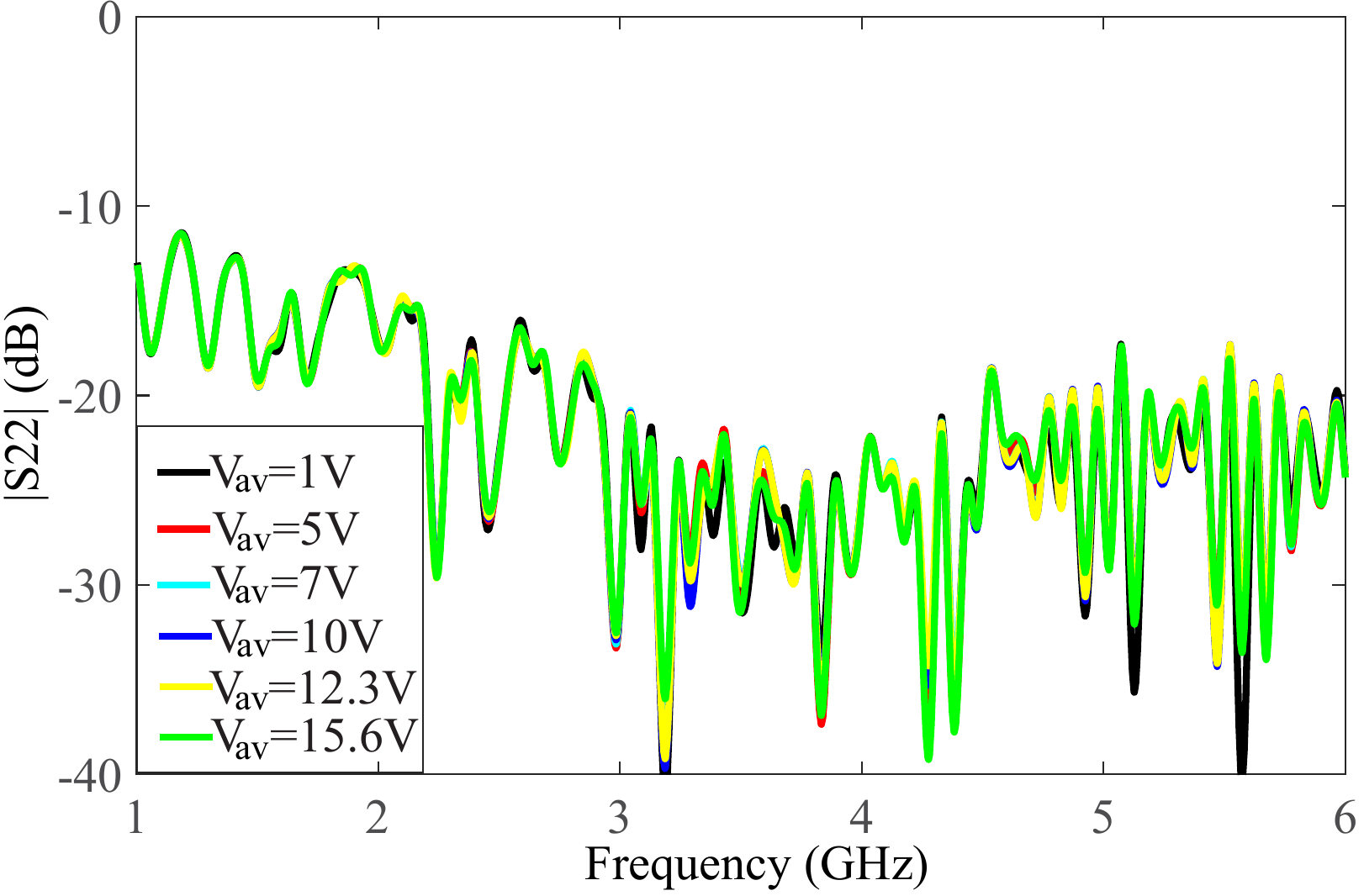}}
		\caption{Experimental scattering parameters of the non-modulated metasurface ($\omega_\text{m}=0$) for different voltages corresponding to different $\epsilon_\text{ant}$s. (a)~$|S_{21}|$. (b)~$|S_{11}|$. (c)~$|S_{22}|$.} 
		\label{Fig:sparam}
	\end{center}
\end{figure}

Figures~\ref{fig:s21},~\ref{fig:s11} and~\ref{fig:s22} plot the experimental scattering parameters of the non-modulated metasurface ($\omega_\text{m}=0$) for different voltages corresponding to different $\epsilon_\text{ant}$s. Figure~\ref{fig:s21} shows that there are three major transmissions through the metasurface around 2.3 GHz, 3.3 GHz and 5 GHz. Hence, we examine a frequency conversion from 2.3 GHz to 3.3 GHz (corresponding to the modulation frequency $\omega_\text{m}=1.06$ GHz), and a frequency down-conversion from 5 GHz to 3.3 GHz (corresponding to the modulation frequency $\omega_\text{m}=1.79$ GHz).

\begin{figure}
	\begin{center}
		\includegraphics[width=1\columnwidth]{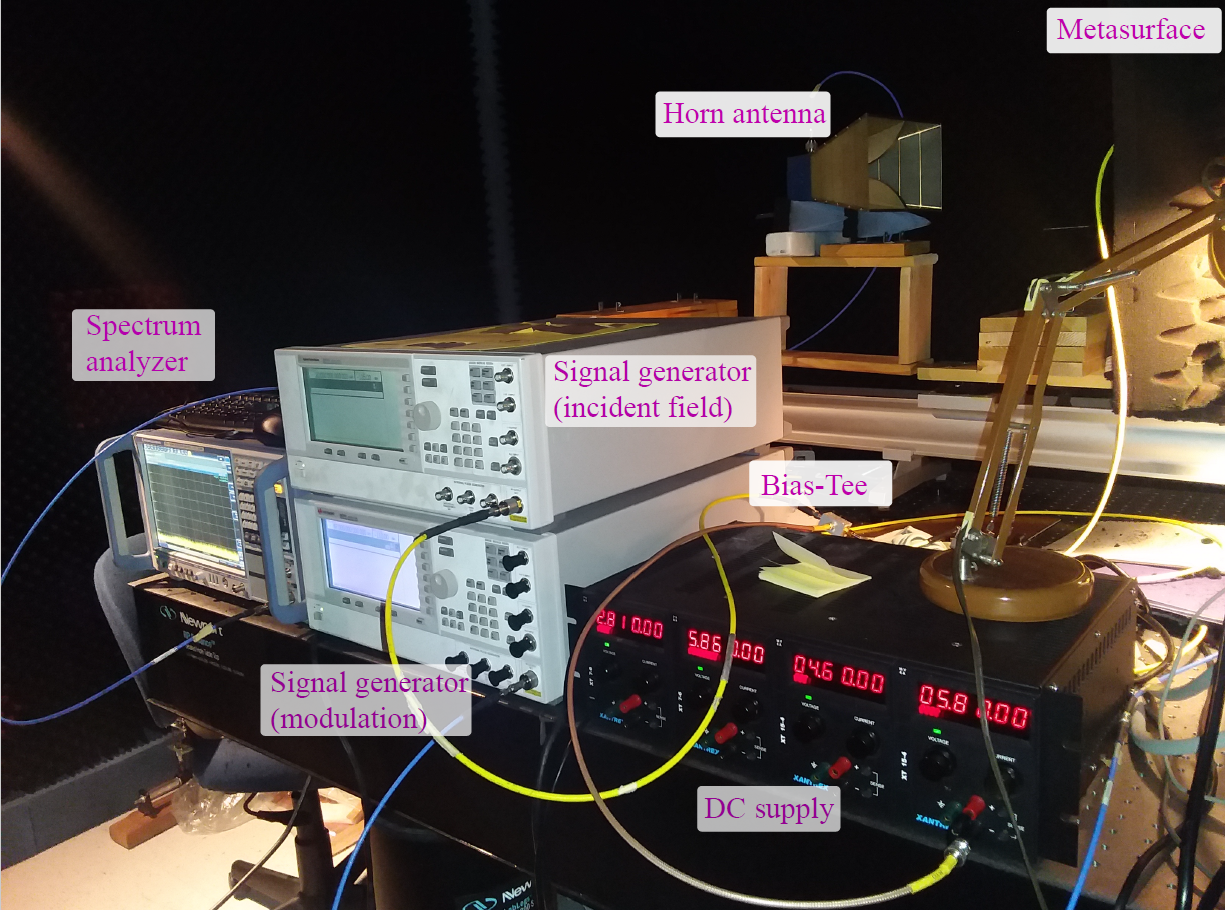}
		\caption{Experimental set-up for measuring the frequency conversion through the time-modulated metasurface.} 
		\label{Fig:spect}
	\end{center}
\end{figure}

We shall stress that the frequency conversion of the metasurface cannot be measured by a vector network analyzer, but using a signal generator and a spectrum analyzer, where a monochromatic wave generated by the signal generator impinges on the metasurface and the transmitted frequency converted wave is measured by a spectrum analyzer. Figures~\ref{Fig:spect} shows the experimental set-up for measuring the frequency conversion through the time-modulated metasurface. The experimental set-up for the measurement of frequency conversion includes two horn antennas, two signal generators, one for the incident signal and the other one for the modulation signal, a spectrum analyzer, a bias-tee for safe integration of the RF modulation bias and the DC bias of varactors, and a DC power supply.

\begin{figure*}
	\begin{center}
		\subfigure[]{\label{fig:inc1} 
			\includegraphics[width=0.87\columnwidth]{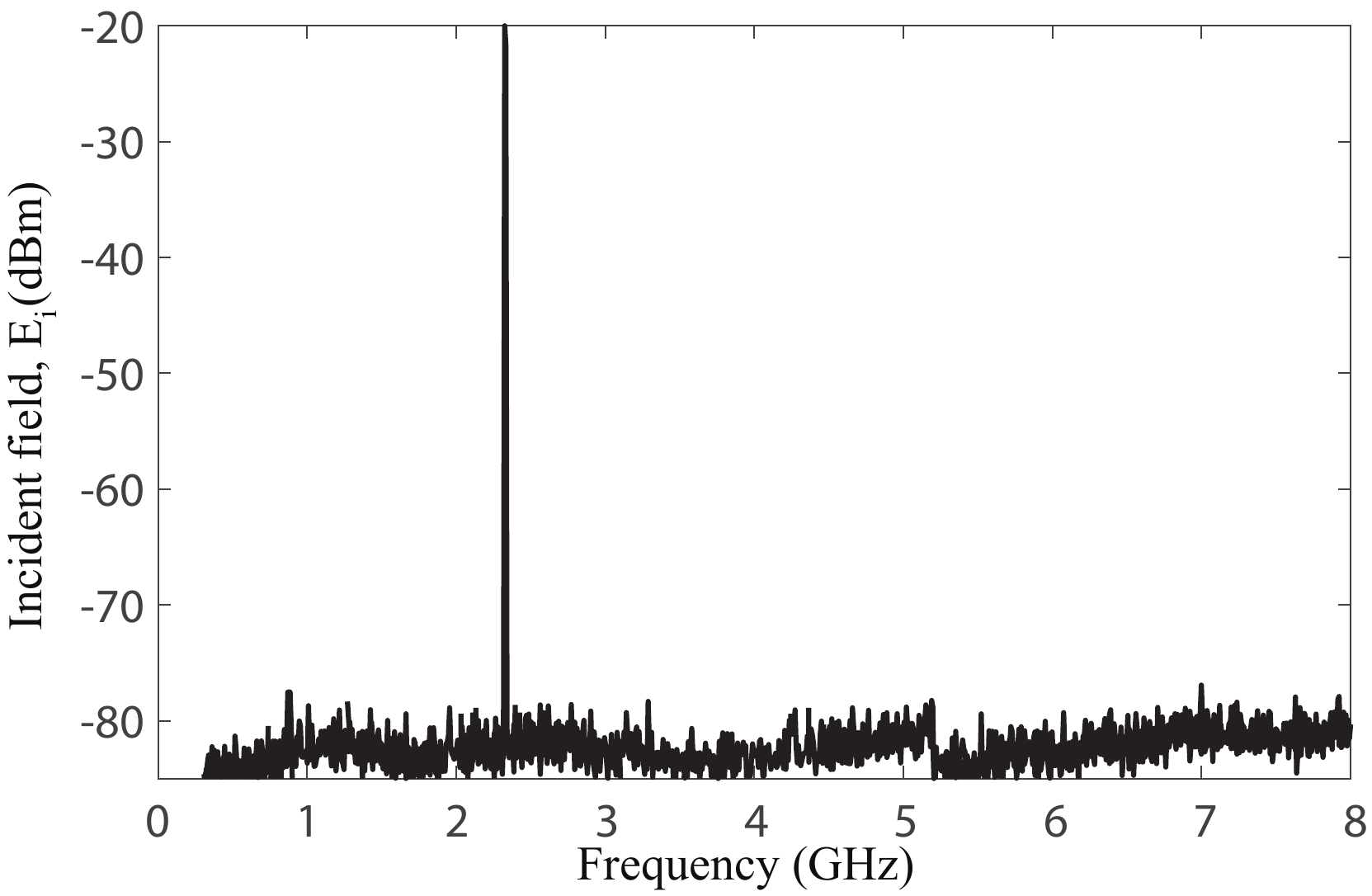}} 
		\subfigure[]{\label{fig:uc} 
			\includegraphics[width=0.87\columnwidth]{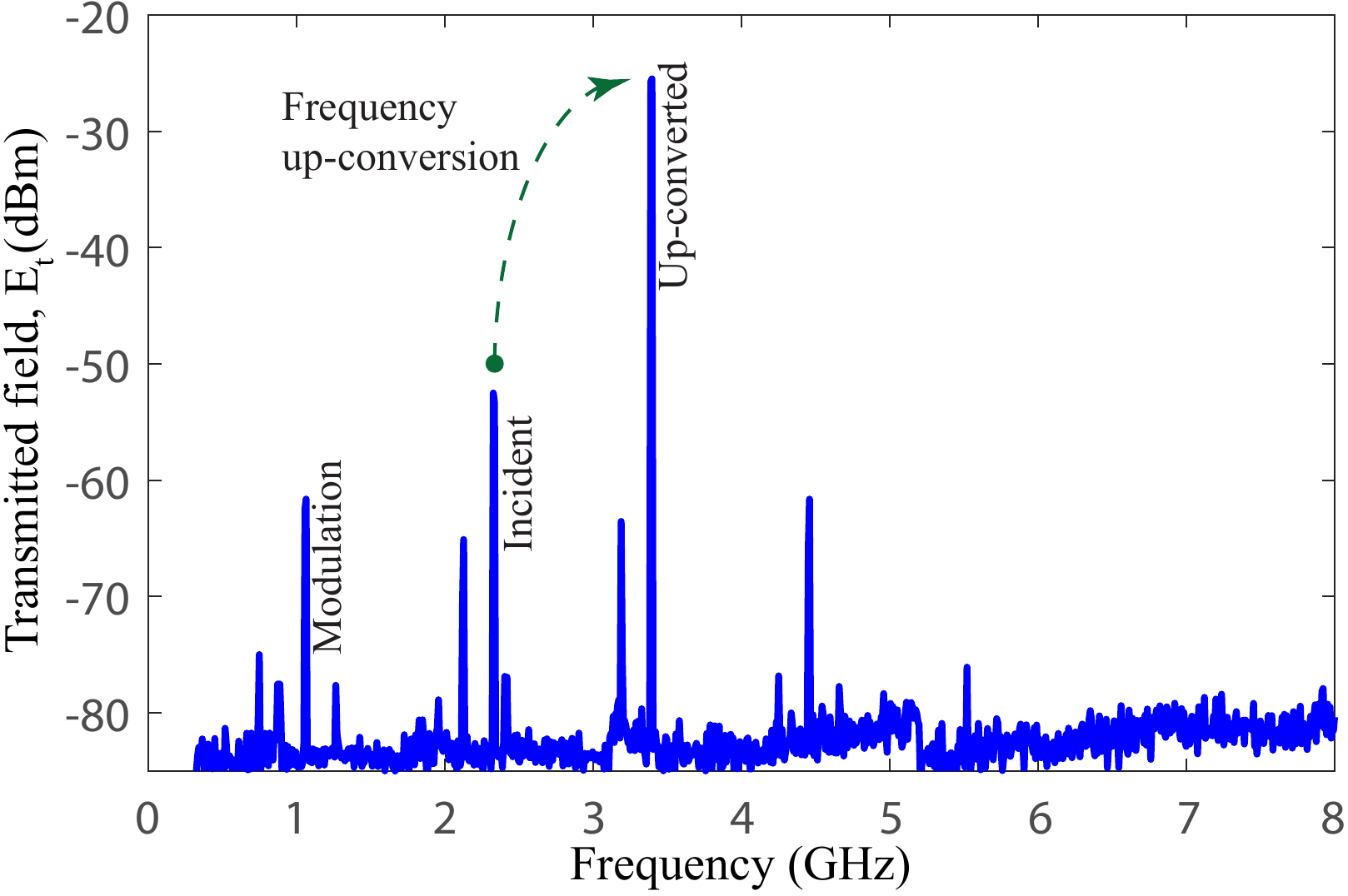}}
		\subfigure[]{\label{fig:inc2} 
			\includegraphics[width=0.87\columnwidth]{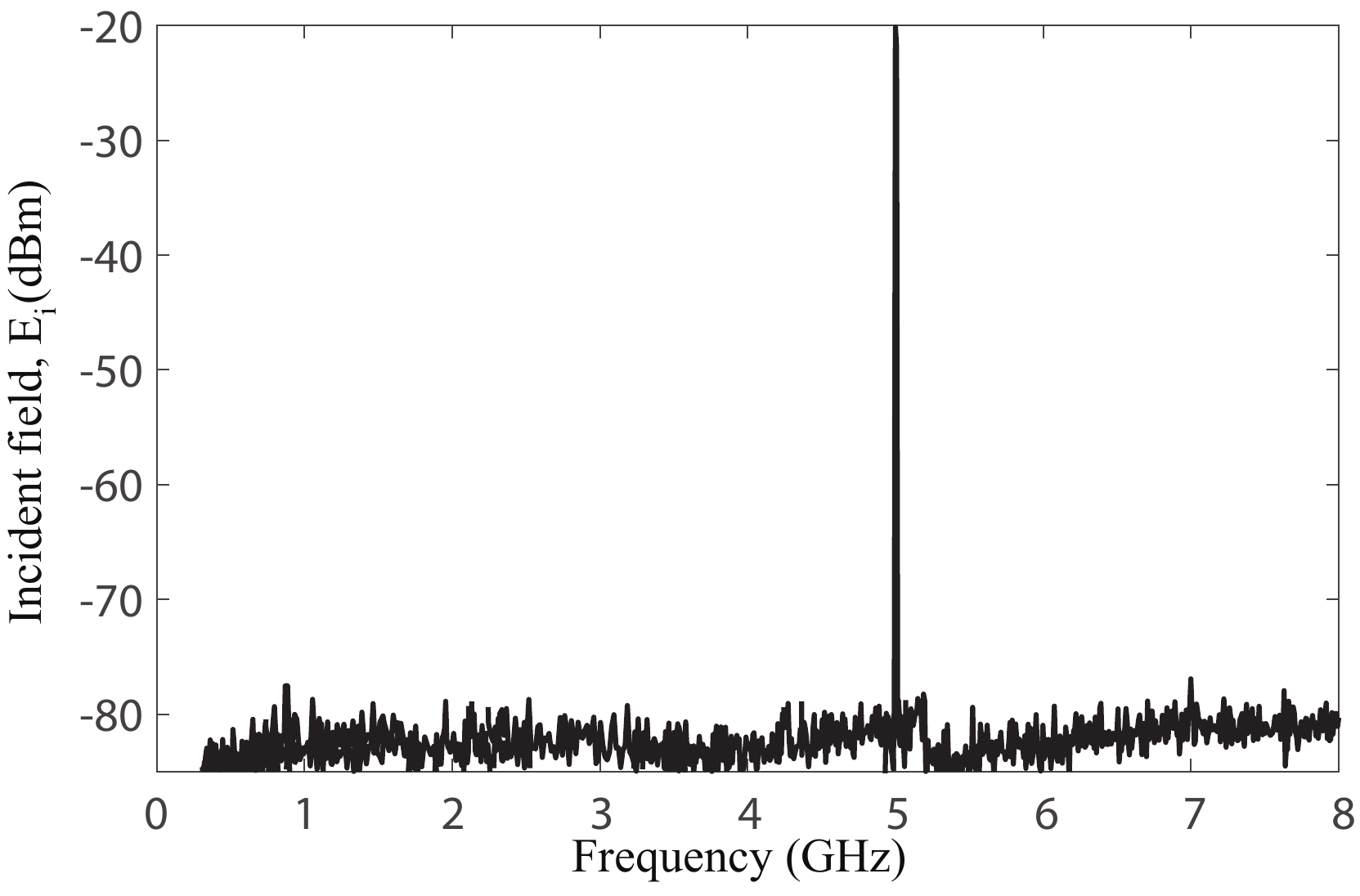}}
		\subfigure[]{\label{fig:dc} 
			\includegraphics[width=0.87\columnwidth]{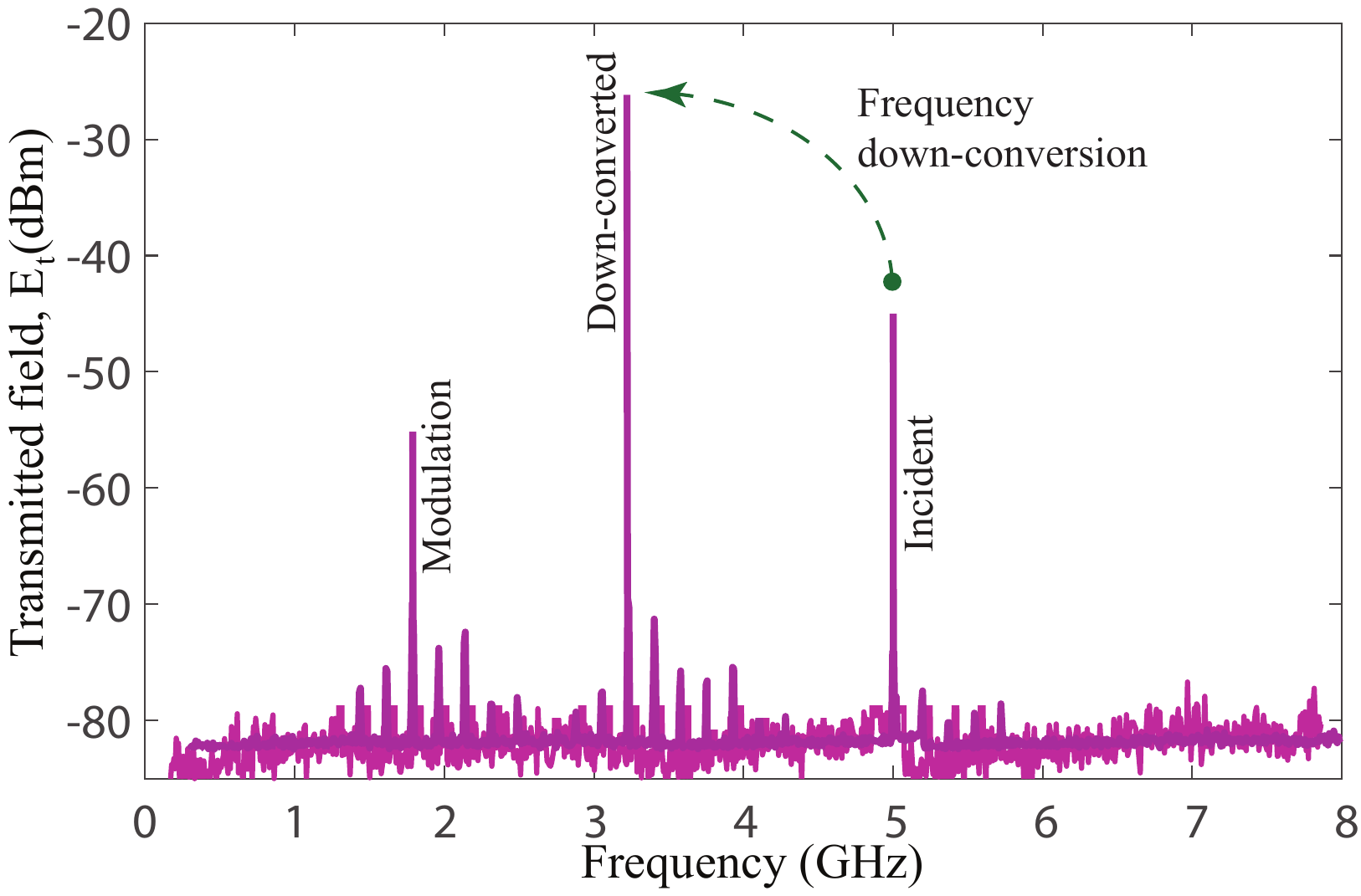}}
		\caption{Experimental results for wave transmission through the frequency converter metasurface. (a) Incident signal at 2.33 GHz. (b) Transmitted up-converted signal at 3.39 GHz (considering modulation frequency of 1.06 GHz). (c) Incident signal at 5 GHz. (d) Transmitted down-converted signal at 3.21 GHz (considering modulation frequency of 1.79 GHz).} 
		\label{Fig:slab_concept}
	\end{center}
\end{figure*}

Figures~\ref{fig:inc1} and~\ref{fig:uc} plot the experimental results for the incident wave to the metasurface and transmitted frequency up-converted wave. In this experiment, the incident signal frequency is at 2.33 GHz, the modulation frequency is at 1.06 GHz, and the transmitted up-converted signal is at 3.39 GHz. It may be seen from Fig.~\ref{fig:uc} that a spurious-free and linear frequency up-conversion is achieved, that is, an up-conversion from 2.33 GHz to 3.39 GHz. The undesired mixing products are suppressed more than 36.3 dB, and the incident wave is suppressed more than 27.06 dB. 

\begin{figure}
	\begin{center} 
		\subfigure[]{\label{Fig:linearity_u}
			\includegraphics[width=0.48\columnwidth]{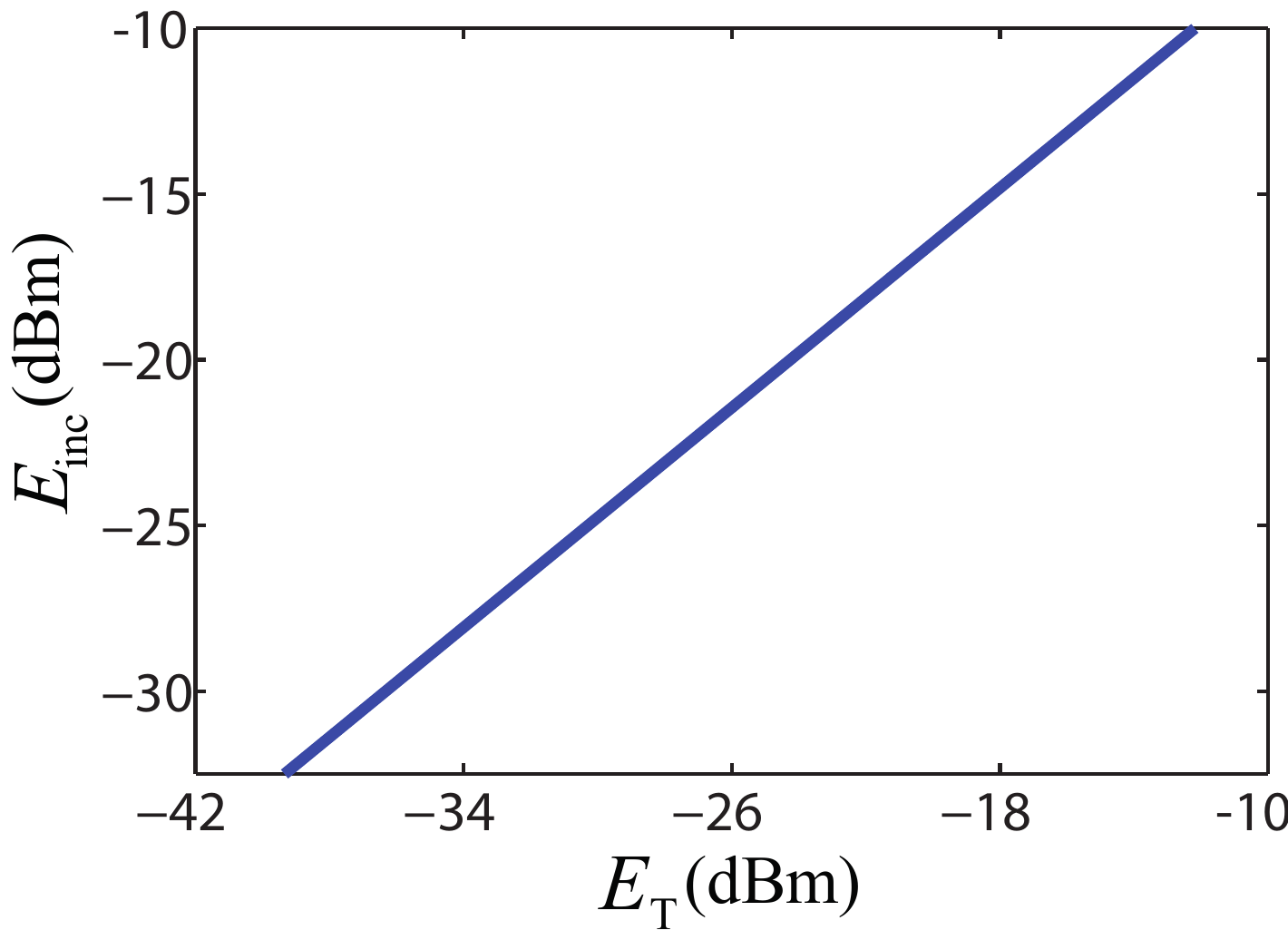}}
		\subfigure[]{\label{Fig:linearity_d}
			\includegraphics[width=0.48\columnwidth]{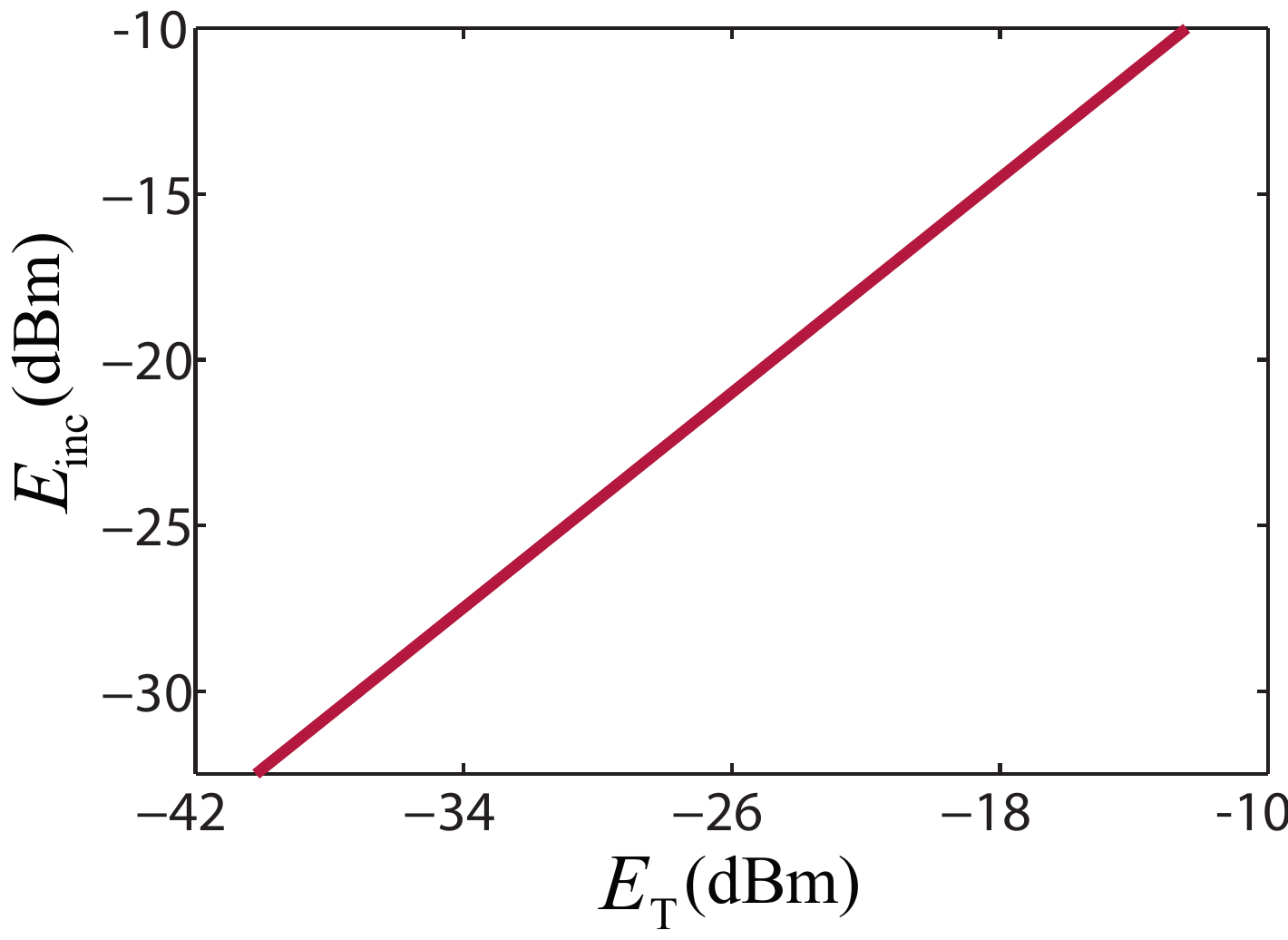}}
		\caption{Experimental results showing the linearity of the temporal frequency converter metasurface for (a) Frequency up-conversion in Fig.~\ref{fig:uc} and (b) Frequency down-conversion in Fig.~\ref{fig:dc}.}
	\end{center}
	\label{Fig:exp_res}
\end{figure}

Figures~\ref{fig:inc2} and~\ref{fig:dc} plot the experimental results for the incident wave to the metasurface and the transmitted frequency down-converted wave. In this experiment, the incident signal frequency is at 5 GHz, the modulation frequency is at 1.79 GHz, and the transmitted up-converted signal is at 3.21 GHz. Figure~\ref{fig:dc} shows that a spurious-free and linear frequency down-conversion is achieved, i.e., a down-conversion from 5 GHz to 3.21 GHz. The undesired mixing products are suppressed more than 28.6 dB, and the incident wave is suppressed more than 18.1 dB. Figure~\ref{Fig:linearity_u} and ~\ref{Fig:linearity_d} demonstrate linear response of the temporal metasurface for up- and down-conversions, respectively. These figures show that the frequency converted transmitted field $E_\text{T}$ linearly follows the incident field $E_\text{inc}$.

\section{Conclusions}

We have proposed the first spurious-free and linear frequency converter metasurface. Our new approach to achieve metasurface-based frequency conversion transmissive temporally modulated supercells was presented here on the basis of surface-interconnector-phase-surface (SIPS) architectures with specifically tailored passbands and stopbands. The proposed form of modulation removes the periodicity of the time modulation, which, combined with suitably created dispersion engineering, realizes spurious-free and linear frequency conversion through a thin sheet. The proposed frequency converter metasurface is capable of implementing large frequency up- and -down conversion ratios. The proposed frequency converter metasurface is very practical, as in real-scenario wireless telecommunication systems, a large frequency conversion is required, i.e., a frequency conversion from an intermediate frequency (VHF/UHF) to microwave frequencies in receivers. In contrast, recently proposed time-varying frequency converters suffer from very low frequency conversion ratios (up-/down-converted frequency is very closed to the input frequency)~\cite{Taravati_APS_2015,Taravati_LWA_2017,Grbic2019serrodyne}.
	
In contrast to conventional nonlinear mixers, the proposed frequency converter temporal metasurface introduces a linear response, where the magnitude of the output frequency-converted signal follows the input signal magnitude. Such a linear response is endowed by the time modulation technique. The proposed frequency converter is very low-profile and formed by a thin (sub-wavelength) metasurface slab which is paramount for practical application.
	
The proposed metasurface inherently provides band-pass filtering, where spurious-free and linear frequency up- and down-conversions occur in a way that the undesired time harmonics are significantly suppressed. The proposed architecture offers extra freedom on controlling the frequency bands as well as the magnitude of the converted frequency, making it an excellent apparatus for versatile wireless communication systems. We have performed a proof-in-principle experiment in the microwave regime for verification. It is worth emphasizing that the proposed theory and architecture are scalable to higher frequencies.	

Furthermore, such a frequency converter metasurface may present conversion gain for greater pumping depths. The magnitude conversion ratio at the output of the frequency converter can be further augmented by appropriate design and fabrication of the architecture. Appropriate elements may be envisioned at terahertz and optics for the realization of time modulation, e.g., using dielectric slabs doped to create p-i-n junction schemes responding to a modulation wave and operating as voltage-controllable capacitors~\cite{Khilo_2011,Fan_PRL_109_2012}.
The proposed concept and technology opens pathways towards several
frequency conversion microwave and nanophotonic components, without requiring bulky antennas and nonlinear mixers, for a variety of applications ranging from telecommunication and biomedical systems to radio astronomy and military radars.

\vspace{7mm}
\hspace{14mm}\textbf{ACKNOWLEDGMENT}

\vspace{1mm}
This work is supported by the Natural Sciences and Engineering Research Council of Canada (NSERC).

\bibliography{Taravati_Reference}

\end{document}